\documentclass[floatfix,showpacs,tightenlines,pra,twocolumn,final,superscriptaddress]{revtex4}%
\usepackage{simplewick}
\usepackage{amsfonts}
\usepackage{amsmath}
\usepackage{amssymb}
\usepackage{graphicx}
\providecommand{\U}[1]{\protect\rule{.1in}{.1in}}
\providecommand{\U}[1]{\protect\rule{.1in}{.1in}}
\providecommand{\U}[1]{\protect\rule{.1in}{.1in}}
\providecommand{\U}[1]{\protect\rule{.1in}{.1in}}
\begin{document}
\title{Effective renormalized multi-body interactions of harmonically confined
ultracold neutral bosons}
\author{P.R. Johnson}
\email{pjohnson@american.edu}
\affiliation{Department of Physics, American University, Washington DC 20016, USA}
\author{D. Blume}
\affiliation{Department of Physics and Astronomy, Washington State University, Pullman,
Washington 99164-2814, USA}
\author{X. Y. Yin}
\affiliation{Department of Physics and Astronomy, Washington State University, Pullman,
Washington 99164-2814, USA}
\author{W.F. Flynn}
\affiliation{Department of Physics, American University, Washington DC 20016, USA}
\author{E. Tiesinga}
\affiliation{Joint Quantum Institute, National Institute of Standards and Technology and
University of Maryland, Gaithersburg, Maryland 20899, USA}
\keywords{Ultracold Atoms, Optical Lattices, Renormalized Perturbation Theory,
Multi-Body Interactions}
\pacs{31.15.ac,31.15.xp,05.30.Jp,67.85.-d}

\begin{abstract}
We calculate the renormalized effective two-, three-, and four-body
interactions for $N$ neutral ultracold bosons in the ground state of an
isotropic harmonic trap, assuming two-body interactions modeled with the
combination of a zero-range and energy-dependent pseudopotential. We work to
third-order in the scattering length $a_{\text{t}}(0)$ defined at zero
collision energy, which is necessary to obtain both the leading-order
effective four-body interaction and consistently include finite-range
corrections for realistic two-body interactions. The leading-order, effective
three- and four-body interaction energies are $U_{3}\left(  \omega\right)
=-(0.85576...)[a_{\text{t}}(0)/\sigma(\omega)]^{2}+2.7921(1)[a_{\text{t}%
}(0)/\sigma(\omega)]^{3}+\mathcal{O}(a_{\text{t}}^{4})$ and $U_{4}%
(\omega)=+(2.43317...)[a_{\text{t}}(0)/\sigma(\omega)]^{3}+\mathcal{O}%
(a_{\text{t}}^{4})$, where $\omega$ and $\sigma(\omega)$ are the harmonic
oscillator frequency and length, respectively, and energies are in units of
$\hbar\omega$. The one-standard deviation error $\pm0.0001$ for the
third-order coefficient in $U_{3}(\omega)$ is due to numerical uncertainty in
estimating a slowly converging sum; the other two coefficients are either
analytically or numerically exact. The effective three- and four-body
interactions can play an important role in the dynamics of tightly confined
and strongly correlated systems. We also performed numerical simulations for a
finite-range boson--boson potential, and it was comparison to the zero-range
predictions which revealed that finite-range effects must be taken into
account for a realistic third-order treatment. In particular, we show that the
energy-dependent pseudopotential accurately captures, through third order, the
finite-range physics, and in combination with the multi-body effective
interactions gives excellent agreement with the numerical simulations,
validating our theoretical analysis and predictions.

\end{abstract}
\date{\today }
\maketitle

\section{Introduction\label{sec:introduction}}

Effective multi-body interactions arise when quantum fluctuations dress the
intrinsic interactions between particles. They play a central role in quantum
field theories and exemplify the significant difference between interactions
in classical and quantum theories. For example, even for a quantum field that
has only intrinsic two-body interactions at high energies, at low-energy
scales, after the high-energy degrees of freedom are coarse-grained away, the
field will manifest at some level effective $n$-body interactions. The ability
to trap and control systems of ultracold neutral atoms
\cite{Bloch2008,Chin2010} has created new opportunities to study this physics
in the laboratory. Effective three-body interactions in the limit of large
two-body scattering length have in particular received a great deal of
attention, motivated both by the predictions of universal behaviors
\cite{Efimov1970,Nielsen1999,Esry1999,Bedaque2000,Kraemer2006,Braaten2007,Braaten2006}
and the ability to use ultracold atoms to study physics ranging from molecular
\cite{Torrontegui2011} to nuclear scales \cite{Beane2007,Meade2009}. Recently,
attention has focused on Efimov-like states and universal behaviors for
four-body systems, again in the limit of large scattering lengths
\cite{Hammer2007,Stecher2009,Pollack2009,Ferlaino2009}.

Here, we focus on the opposite regime of weakly interacting neutral bosons
with small scattering lengths. Even in this limit, effective higher-body
interactions can be important, particularly for tightly confined or strongly
correlated particles. This is seen dramatically in \cite{Will2010}, where a
superfluid of bosonic atoms is quenched by suddenly increasing the depth of an
optical lattice. After the quench, which creates a non-equilibrium state of
strongly correlated bosons, beating effects due to multiple distinct
interaction energies, as expected from effective three- and higher-body
interactions \cite{Johnson2009,Tiesinga2011}, are seen in the collapse and
revival oscillations of the first-order coherence. Effective multi-body
interactions should also have played a role in previous collapse and revival
experiments \cite{Greiner2002b,Anderlini2006,Sebby2007}, although in those
cases inhomogeneities may have masked their signature. More recently,
effective three- and four-body interactions have been used to demonstrate
atom-number sensitive photon-assisted tunneling in optical lattices
\cite{Ma2011}, and their influence has been seen in precision measurements on
Mott-insulator states of ultracold atoms \cite{Mark2011}. A number of studies
also suggest that elastic multi-body interactions can play an interesting role
in generating exotic quantum phases in optical lattices or modifying the
superfluid to Mott-insulator phase transition
\cite{Buchler2007,Chen2008,Schmidt2008,Sansone2009,Mazza2010,Zhou2010,Will2011,Singh2012}%
.

In this paper, we use renormalized quantum field theory \cite{Srednicki2007}
to calculate the perturbative ground-state energy for $N$ ultracold neutral
bosons in a three dimensional isotropic harmonic potential with angular
frequency $\omega,$ and extract from it the effective $m$-body interaction
energies $U_{2}(\omega),$ $U_{3}(\omega),$ and $U_{4}(\omega)$ as a function
of $\omega.$ The key purpose of the present paper is to (i) systematically
develop a renormalized quantum field theory approach for ultracold trapped
bosons including finite-range effects, (ii) determine the leading-order
four-body interaction, and (iii) validate the formalism through comparison
with numerical results. To obtain effective four-body interaction energies it
is necessary to work through third order in the two-body scattering length. We
use renormalized perturbation theory (see~\cite{Srednicki2007}), which
develops an expansion around \emph{physical} as opposed to \emph{bare}%
\ coupling parameters, to systematically cancel the multiple divergences that
arise at higher-orders in quantum field perturbation theory. (In this paper,
the physical coupling parameter is defined in terms of the measured scattering
length, or alternatively the measured energy shift, for two interacting
ultracold boson in a harmonic trap at a specified trap frequency.)
Renormalized perturbation theory, which is more commonly used in high-energy
physics, in this context naturally describes how the effective interactions
depend on trap frequency. An example of the power of renormalized perturbation
theory to capture low-energy physics is that we independently reproduce,
through third order, the two-body ground-state energies calculated in
\cite{Busch1998}. More fundamentally, the analysis in this paper provides an
explicit example of renormalization physics and running coupling constants
that can be directly probed using trapped ultracold bosonic atoms, and used to
test central concepts in effective field theory.

To calculate effective interactions for confined bosons, we first assumed that
the two-body interactions could be described in the low-energy, $s$-wave limit
by an \emph{energy-independent zero-range} $\delta$-function pseudopotential.
To test our perturbative predictions, we then numerically calculated $N$-boson
ground-state energies using a \emph{finite-range} two-body Gaussian model
potential. Comparison with the numerical results revealed that finite-range
effects must also be taken into account for an accurate description of
realistically interacting bosons. In this paper, we show that both the finite
range effects and effective interactions are accurately captured by the
combination of zero-range and \emph{energy-dependent} $\delta$-function
pseudopotentials. Including the finite-range corrections, we are able to
validate our analytic and numerical calculations of all perturbation theory
coefficients through third order.

\begin{figure}[ptb]
\centering{\includegraphics[]{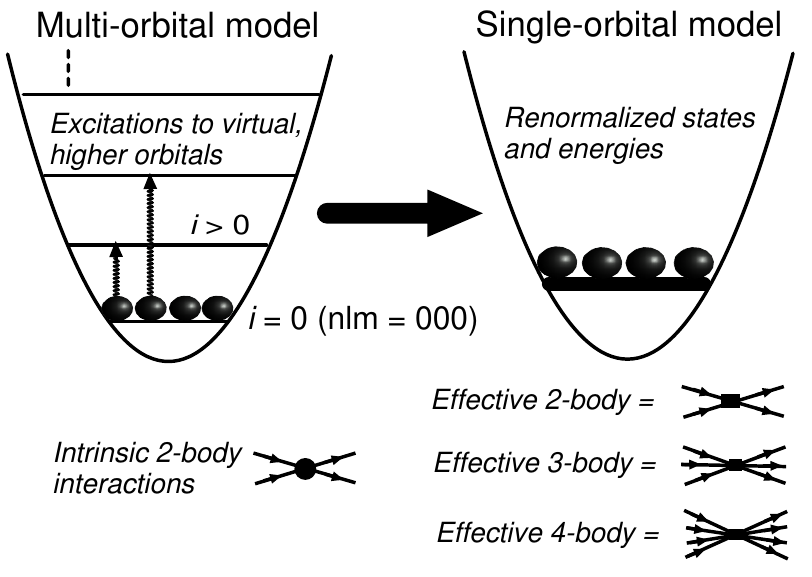}} \caption{Illustration of
the idea of replacing a multiple-orbital (or multiple-band) model with
orbitals $i=0,1,...$ with only intrinsic two-body interactions by a
single-orbital (or single-band) model with effective multi-body interactions
between renormalized states. A ground state multi-body model can be useful
when virtual excitations of bosons to excited vibrational levels are
important.}%
\label{fig:harmonicwellcartoon}%
\end{figure}

The basic idea in our approach is the following: we \textquotedblleft
integrate out\textquotedblright\ excited vibrational states thereby trading a
multi-orbital theory with intrinsic two-body interactions for a single-orbital
theory with effective multi-body interactions. The latter can provide a simple
but powerful alternative description of the low-energy few-body physics. The
quantum fluctuations to excited states both dress the two-body interactions
and generate effective higher-body interactions. The idea is illustrated in
Fig.~\ref{fig:harmonicwellcartoon}. We showed in
\cite{Johnson2009,Tiesinga2011} how this approach can be used to approximately
incorporate the influence of higher bands via the simple modification of
adding higher-body interactions to the single-band Bose-Hubbard model
\cite{Fisher1989,Jaksch1998}.

Beyond applying directly to ultracold neutral bosons in an isotropic harmonic
potential, our results can give qualitative insight into the effective
interactions for other trapping potentials. They can also be used for rough
approximations to the effective two-, three-, and four-body interactions in
anisotropic potentials, and for neutral bosons in optical lattices. In the
latter case, however, anharmonicities are important. For example, we estimate
an approximately 30\% anharmonic correction to the three-body interactions for
$^{87}$Rb in typical lattices. The role of anharmonicities for
collapse-and-revival dynamics in optical lattice systems has been analyzed
further in \cite{Mering2011}. Inhomogeneities and the effect of a background
harmonic potential on lattice collapse-and-revival dynamics has been studied
in \cite{Schachenmayer2011,Buchhold2011}.

Tunneling also has an influence on collapse and revival in optical lattices
\cite{Rigol2006,Fischer2008,Wolf2010}. In deep (post-quench) lattices the
typical tunneling energy is nearly an order of magnitude smaller than the
effective three-body interaction energy, making the latter effect dominant.
Tunneling should, however, be of comparable importance to the effective
four-body interactions. Approaches applying effective interaction methods to
tunneling in lattice or multi-well systems include
\cite{Ananikian2006,Folling2007,Zhou2011,Pielawa2011}, and related methods for
analyzing physics involving interactions, correlations, higher bands, and
quantum tunneling in lattice systems include
\cite{Alon2005,Alon2007,Hazzard2010,Luhmann2012,Bissbort2011,Cao2011}.
Fermionic systems and fermion-boson mixtures also yield interesting types of
effective interactions that have received increasing attention (e.g.,
\cite{Blume2007,Luhmann2008,Lutchyn2009,Mering2011,Rotureau2010,Will2011}), as
well as three-body interactions of fermions and polar molecules in lattices
\cite{Buchler2007}.

For experiments with $^{87}$Rb at typical lattice densities the recombination
rate \cite{Burt1997,Esry1999} is one or more orders of magnitude smaller than
the frequencies associated with both the effective three- and four-body
energies, and therefore the elastic effective interactions described in the
present paper are more important than inelastic multi-body interactions
driving loss. Roughly, we expect three-body recombination to scale at fourth
order in the scattering length \cite{Fedichev1996}, and in the future we would
like to understand both elastic and inelastic interactions in a unified
framework. The role of effective three-body interactions in thermalizing a
homogenous 1D Bose gas has also been studied \cite{Mazets2010}, and it would
be interesting to investigate this physics in the context of a 3D optical
lattice system.

The remainder of this paper is organized as follows. In
Sec.~\ref{sec:overview}, we provide an overview of our results.
Section~\ref{sec:comparisonofenergies} compares the perturbation theory
predictions to numerical estimates for finite-range interactions.
Sections~\ref{sec:1stand2ndordereffectiveinteractions} and
\ref{sec:thirdorderenergies} describe the details of the renormalized
perturbation theory used to obtain the effective multi-body interactions.
Section~\ref{sec:1stand2ndordereffectiveinteractions} defines the renormalized
Hamiltonian and derives the first- and second-order corrections, while
Sec.~\ref{sec:thirdorderenergies} derives the two-, three-, and four-body
interaction energies through third order. Section~\ref{sec:conclusion}
summarizes our results and conclusions. Finally, the appendices give
derivations of a number of technical results used in the paper.

\section{Overview\label{sec:overview}}

We find the effective interactions of $N$ ultracold bosons in the ground state
of an isotropic harmonic oscillator with pairwise interactions modeled by a
zero-range $\delta$-function pseudopotential
\begin{equation}
\mathcal{V}_{2}(\mathbf{r}_{i}-\mathbf{r}_{j})=g_{2}\delta^{(3)}%
(\mathbf{r}_{i}-\mathbf{r}_{j}), \label{deltafunctionpotential}%
\end{equation}
where $\mathbf{r}_{i}$ is the position vector of the $i^{\text{th}}$ boson. We
assume there are no intrinsic three- or higher-body interactions. The two-body
coupling constant $g_{2}$ is related to $a_{\text{t}}(0)$, at first order in
perturbation theory, by $g_{2}=4\pi\left(  \hbar^{2}/m_{\text{A}}\right)
a_{\text{t}}(0)+\mathcal{O}([a_{\text{t}}(0)]^{2}),$ where $m_{\text{A}}$ is
the boson mass, $a_{\text{t}}(0)$ is the physical $s$-wave scattering length
measured in the limit that the trap frequency and collision energy go to zero,
and $\mathcal{O}([a_{\text{t}}(0)]^{2})$ are terms of order $[a_{\text{t}%
}(0)]^{2}$ and higher. At higher orders, the relationship between $g_{2}$ and
$a_{\text{t}}(0)$ is modified, and in
Secs.~\ref{sec:1stand2ndordereffectiveinteractions} and
\ref{sec:thirdorderenergies} we generalize the perturbation theory as an
expansion around the physical \emph{trap} scattering length $a_{\text{t}%
}(\omega_{0})$ defined for a harmonic potential with frequency $\omega_{0}.$
In this overview, we summarize our results to third order in $a_{\text{t}%
}(0),$ i.e., the special case $\omega_{0}=0.$

We obtain the ground-state energy of $N$ bosons as an expansion $E=\sum
_{n=0}E^{(n)},$ where $E^{(n)}$ is proportional to $[a_{\text{t}}(0)]^{n}.$
Throughout this paper energies are expressed in units of the harmonic
oscillator energy $\hbar\omega.$ The zeroth-order (one-body) energy is
$E^{(0)}(\omega)=\varepsilon_{0}N,$ where $\varepsilon_{0}=3/2$ is the
dimensionless single-particle ground-state energy. The $n^{\text{th}}$-order
energies for $n>0$ can be expanded as
\begin{equation}
E^{(n)}=\sum_{m=2}\binom{N}{m}U_{m}^{(n)}(\omega), \label{EnN}%
\end{equation}
where $\binom{N}{m}$ is the binomial coefficient. The sum goes up to the
minimum of $N$ and $n+1$, and the $n^{\text{th}}$-order contributions to the
$m$-body interaction energies (in units of $\hbar\omega)$ are
\begin{equation}
U_{m}^{(n)}(\omega)=c_{m}^{(n)}\left(  \frac{a_{\text{t}}(0)}{\sigma(\omega
)}\right)  ^{n}, \label{Unm}%
\end{equation}
where the harmonic oscillator length for an isotropic potential with frequency
$\omega$ is%
\begin{equation}
\sigma(\omega)=\sqrt{\hbar/m_{\text{A}}\omega}.
\end{equation}

Table~\ref{Table:CoefficientsCnm} gives the values of $c_{m}^{(n)}$ obtained
in Secs.~\ref{sec:1stand2ndordereffectiveinteractions}
and~\ref{sec:thirdorderenergies}. The two-body coefficients $c_{2}^{(1)}%
,c_{2}^{(2)}, $ and $c_{2}^{(3)}$ independently reproduce the results in
\cite{Busch1998}, if the exact solution found there is expanded through third
order. The coefficient $c_{2}^{(3)}$, in particular, is nontrivial and
provides a strong consistency check that the renormalized perturbation theory
captures the two-body low-energy interactions correctly.

The analytic value of the three-body coefficient $c_{3}^{(2)}$ was previously
found in \cite{Johnson2009}. The coefficient $c_{3}^{(3)}$ found here extends
that result to third order in $a_{\text{t}}(0)$. The value of $c_{3}^{(3)}$
given in Table~\ref{Table:CoefficientsCnm} combines both analytic and
approximate numerical results, and the uncertainty is due to the slow
convergence of one of the numerically determined sums (see
App.~\ref{app:alpha3(33)}).

We also obtain the coefficient $c_{4}^{(3)},$ which gives the leading-order
contribution to the effective four-body energy. The coefficient $c_{4}^{(3)}$
combines numerical and analytic results, but unlike $c_{3}^{(3)}$ has high
precision because of the fast convergence of all the contributing terms. Note
that $c_{3}^{(3)}$ and $c_{4}^{(3)}$ have similar magnitudes, and consequently
we need to include the effective three-body corrections when effective
four-body effects are important or of interest. At the end of
Sec.~\ref{sec:overview}, we show that the correction from the third-order
terms becomes significant for ultracold atoms in trap potentials with
relatively tight confinement. The coefficients $c_{3}^{(3)}$ and $c_{4}^{(3)}$
have not previously been reported in the literature.

\begin{table}[t]
{\footnotesize \centering$%
\begin{tabular}
[c]{l}%
Effective Interaction Energy Coefficients\\\hline\hline
\ \ \ \ \emph{Two-body}\\
$c_{2}^{(1)}=(2/\pi)^{1/2}=+0.79788...$\\
$c_{2}^{(2)}=(2/\pi)(1-\log2)=+0.19535...$\\
$c_{2}^{(3)}=(2/\pi)^{3/2}(1-\frac{\pi^{2}}{24}-3\log2+\frac{3}{2}\log
^{2}2)=-0.39112...$\\
$d_{2}^{(1,2)}=(3/4)(2/\pi)^{1/2}=+0.59841...$\\
\ \ \ \ \emph{Three-body}\\
$c_{3}^{(2)}=(2/\pi)\{-4\sqrt{3}+6[1-2\log2-\log(2-\sqrt{3})]\}$\\
$\ \ \ \ \ \ =-0.85576...$\\
$c_{3}^{(3)}=-12(2/\pi)^{1/2}(1-\log2)\alpha_{3}^{(2)}+12\alpha_{3}%
^{(3)}-6\alpha_{4,3}^{(3)}-18\alpha_{5}^{(3)}$\\
$\ \ \ \ \ \ =+2.7921(1)$\\
\ \ \ \ \emph{Four-body}\\
$c_{4}^{(3)}=48\alpha_{4,1}^{(3)}+48\alpha_{4,2}^{(3)}-72\alpha_{5}%
^{(3)}=+2.43317...$%
\end{tabular}
\ \ \ \ \ \ \ \ \ \ \ \ \ \ \ $ }\caption{The coefficients $c_{m}^{(n)}$ and
$d_{2}^{(1,2)}$, which give the $n^{\text{th}}$-order correction to the
$m$-body effective interaction energies $U_{m}^{(n)}\left(  \omega\right)  $
[see Eqs. (\ref{Unm}) and (\ref{U2eff12})] for neutral bosons in an isotropic
harmonic potential. The results for $c_{2}^{(1)},c_{2}^{(2)},c_{2}^{(3)}%
,d_{2}^{(1,2)},$ and $c_{3}^{(2)}$ are exact. The coefficients $c_{3}^{(3)}$
and $c_{4}^{(3)}$ are given in terms of parameters $a_{3}^{(2)},\alpha
_{3}^{(3)},$ etc., defined in Table~\ref{table:alphaandbeta}. We have obtained
exact analytic expressions for $\alpha_{3}^{(2)},\alpha_{4,3}^{(3)},$ and
$\alpha_{5}^{(3)}.$ The numerical approximations for $\alpha_{4,1}^{(3)}$ and
$\alpha_{4,2}^{(3)}$ are obtained to very high precision, but slow convergence
of the expression giving $\alpha_{3}^{(3)}$ is responsible for the uncertainty
in the value of $c_{3}^{(3)}.$}%
\label{Table:CoefficientsCnm}%
\end{table}

In Sec.~\ref{sec:comparisonofenergies}, we compare the predictions for
zero-range interactions to numerical calculations for a Gaussian boson-boson
interaction potential and find significant effects from its finite-range
nature. We show that these are accurately modeled by adding to the zero-range
pseudopotential $\mathcal{V}_{2}$ an\emph{\ energy-dependent}
(higher-derivative) pseudopotential \cite{Beane2007}
\begin{equation}
\mathcal{V}_{2}^{\prime}(\mathbf{r}_{i}-\mathbf{r}_{j})=-\frac{g_{2}^{\prime}%
}{2}[\overleftarrow{\nabla}_{ij}^{2}\delta^{(3)}(\mathbf{r}_{i}-\mathbf{r}%
_{j})+\delta^{(3)}(\mathbf{r}_{i}-\mathbf{r}_{j})\overrightarrow{\nabla}%
_{ij}^{2}],\label{V2prime}%
\end{equation}
which has been symmetrized to make it Hermitian. The operators $\overleftarrow
{\nabla}_{ij}$ and $\overrightarrow{\nabla}_{ij}$ are gradients with respect
to the relative separation $\mathbf{r}_{i}-\mathbf{r}_{j},$ acting to the left
and right, respectively. The coupling constant is
\begin{equation}
g_{2}^{\prime}=\left(  4\pi\frac{\hbar^{2}}{m_{\text{A}}}\right)  \left(
\frac{1}{2}r_{\text{eff}}[a_{\text{t}}(0)]^{2}\right)  +\mathcal{O}%
(r_{\text{eff}}[a_{\text{t}}(0)]^{3}),\label{g2prime}%
\end{equation}
where $r_{\text{eff}}$ is the effective range \cite{Taylor2002}. To
first-order in $g_{2}^{\prime},$ the shift to the $N$-body ground-state energy
is
\begin{equation}
E^{(1,2)}=\binom{N}{2}U_{2}^{(1,2)}(\omega)\label{E12}%
\end{equation}
with%
\begin{equation}
U_{2}^{(1,2)}(\omega)=d_{2}^{(1,2)}\left(  \frac{r_{\text{eff}}}{\sigma
(\omega)}\right)  \left(  \frac{a_{\text{t}}(0)}{\sigma(\omega)}\right)
^{2}.\label{U2eff12}%
\end{equation}
The superscript $(1,2)$ indicates that the term is first order in
$r_{\text{eff}}$ and second order in $a_{\text{t}}(0)$, and $d_{2}^{(1,2)}%
\ $is given in Table~\ref{Table:CoefficientsCnm}.

The potential $\mathcal{V}_{2}^{\prime}$ is proportional to $r_{\text{eff}%
}[a_{\text{t}}(0)]^{2}/\sigma(\omega)^{3}$ and we consider in this paper a
regime where $a_{\text{t}}(0)\approx r_{\text{eff}}\ll\sigma(\omega),$ such
that $\mathcal{V}_{2}^{\prime}$ and therefore $U_{2}^{(1,2)}(\omega)$ can be
treated as if the contribution is third order in $a_{\text{t}}(0)$. This
approach is supported by the comparison between the perturbative energies and
the energies for the Gaussian potential with spatial widths $r_{0}%
\lesssim0.01\sigma(\omega)$ in Sec.~\ref{sec:comparisonofenergies}. Adding the
contribution $U_{2}^{(1,2)}(\omega)$ to the two-body interaction energy
extends our results to more realistic systems, like ultracold atoms that
interact through finite-range van der Waals potentials.

Equation~(\ref{EnN}) organizes the $N$-body energy in powers of the free-space
$s$-wave scattering length $a_{\text{t}}(0).$ Alternatively, combining our
results, we can reorganize the energy in terms of $m$-body contributions as

\begin{flalign}
E =\varepsilon_{0}N &+\frac{1}{2!}U_{2}(\omega)N(N-1)\label{Energy}&\\
  &+\frac{1}{3!}U_{3}(\omega)N(N-1)(N-2)\nonumber\\
  &+\frac{1}{4!}U_{4}(\omega)N(N-1)(N-2)(N-3)+...,\nonumber
\end{flalign}

where through third order the two-body interaction energy is

\begin{flalign}
U_{2}(\omega) &  =c_{2}^{(1)}\left(  \frac{a_{\text{t}}(0)}{\sigma(\omega
)}\right)  +c_{2}^{(2)}\left(  \frac{a_{\text{t}}(0)}{\sigma(\omega)}\right)
^{2}+c_{2}^{(3)}\left(  \frac{a_{\text{t}}(0)}{\sigma(\omega)}\right)
^{3}&\nonumber\\
&  +d_{2}^{(1,2)}\left(  \frac{r_{\text{eff}}}{\sigma(\omega)}\right)  \left(
\frac{a_{\text{t}}(0)}{\sigma(\omega)}\right)  ^{2}+\mathcal{O}\left(
\frac{[a_{\text{t}}(0)]^{4}}{[\sigma(\omega)]^{4}}\right)  \nonumber\\
&  +\mathcal{O}\left(  \frac{r_{\text{eff}}[a_{\text{t}}(0)]^{3}}%
{[\sigma(\omega)]^{4}}\right)  ,\label{U2w}%
\end{flalign}

the three-body interaction energy is%

\begin{flalign}
U_{3}(\omega)  & =c_{3}^{(2)}\left(  \frac{a_{\text{t}}(0)}{\sigma(\omega
)}\right)  ^{2}+c_{3}^{(3)}\left(  \frac{a_{\text{t}}(0)}{\sigma(\omega
)}\right)  ^{3}+\mathcal{O}\left(  \frac{[a_{\text{t}}(0)]^{4}}{[\sigma
(\omega)]^{4}}\right) & \nonumber\\
& +\mathcal{O}\left(  \frac{r_{\text{eff}}[a_{\text{t}}(0)]^{3}}%
{[\sigma(\omega)]^{4}}\right)  ,\label{U3w}%
\end{flalign}
and the four-body interaction energy is%
\begin{flalign}
U_{4}(\omega) &=c_{4}^{(3)}\left(  \frac{a_{\text{t}}(0)}{\sigma(\omega
)}\right)  ^{3}+\mathcal{O}\left(  \frac{[a_{\text{t}}(0)]^{4}}{[\sigma
(\omega)]^{4}}\right) &\nonumber\\
&+\mathcal{O}\left(  \frac{r_{\text{eff}}[a_{\text{t}}(0)]^{3}}%
{[\sigma(\omega)]^{4}}\right)  .\label{U4w}&%
\end{flalign}

The four-body interaction energy $U_{4}(\omega)$, although comparatively
small, can lead to qualitatively important effects, particularly for traps
with stronger confinement. For example, for $N=4$ $^{87}$Rb atoms and
$a_{\text{t}}(0)/\sigma(\omega)=0.05,$ corresponding to a $10^{4}$ Hz trap
frequency, the four-body energy should generate a distinct approximately $60$
Hz beating frequency in collapse-and-revival oscillations, using our harmonic
trap results to estimate the energy in an optical lattice potential. These
effects should be measurable as long as tunneling and trap inhomogeneities are
sufficiently reduced \cite{Will2010}.

Using the effective interaction energies in
Eqs.~\eqref{U2w}, \eqref{U3w}, and \eqref{U4w}%
, we can construct a single-orbital effective Hamiltonian
\begin{equation}
H_{\text{eff}}(\omega)=\varepsilon_{0}\hat{a}^{\dag}\hat{a}+\sum_{m=2}\frac
{1}{m!}U_{m}(\omega)\hat{a}^{\dag m}\hat{a}^{m}, \label{Heff}%
\end{equation}
where $\hat{a}$ ($\hat{a}^{\dag}$) annihilates (creates) a boson in a
renormalized single-particle ground state. The effective Hamiltonian can be
used to incorporate some higher-band physics, via effective multi-body
interactions, into a single-band Bose-Hubbard model \cite{Johnson2009}.

The effective interaction energies can be tuned by changing either the
scattering length $a_{\text{t}}(0),$ for example with a Feshbach resonance
\cite{Chin2010}, or the trap frequency $\omega$ of the confinement
\cite{Bolda2002,Blume2002}. For example, for a fixed $a_{\text{t}}(0),$ this
tuning follows from rewriting the $U_{m}(\omega)$ in terms of the
characteristic scattering energy $\hbar\omega_{\text{s}}=\hbar^{2}%
/m_{\text{A}}[a_{\text{t}}(0)]^{2}$. That is, we write $\tilde{U}_{m}%
(\omega)=U_{m}(\omega)(\omega/\omega_{\text{s}})$ such that%
\begin{align}
\tilde{U}_{2}(\omega) &  =c_{2}^{(1)}(\omega/\omega_{\text{s}})^{3/2}%
+c_{2}^{(2)}(\omega/\omega_{\text{s}})^{2}\label{U2tilda}\\
& +(c_{2}^{(3)}+d_{2}^{(1,2)}[r_{\text{eff}}/a_{\text{t}}(0)])(\omega
/\omega_{\text{s}})^{5/2}+\mathcal{O}[(\omega/\omega_{\text{s}})^{3}%
],\nonumber\\
\tilde{U}_{3}(\omega) &  =c_{3}^{(2)}(\omega/\omega_{\text{s}})^{2}%
+c_{3}^{(3)}(\omega/\omega_{\text{s}})^{5/2}+\mathcal{O}[(\omega
/\omega_{\text{s}})^{3}],\label{U3tilda}\\
\tilde{U}_{4}(\omega)&=c_{4}^{(3)}(\omega/\omega_{\text{s}})^{5/2}%
+\mathcal{O}[(\omega/\omega_{\text{s}})^{3}].\label{U4tilda}%
\end{align}

\begin{figure}[ptb]
\centering\includegraphics[width=1\columnwidth]{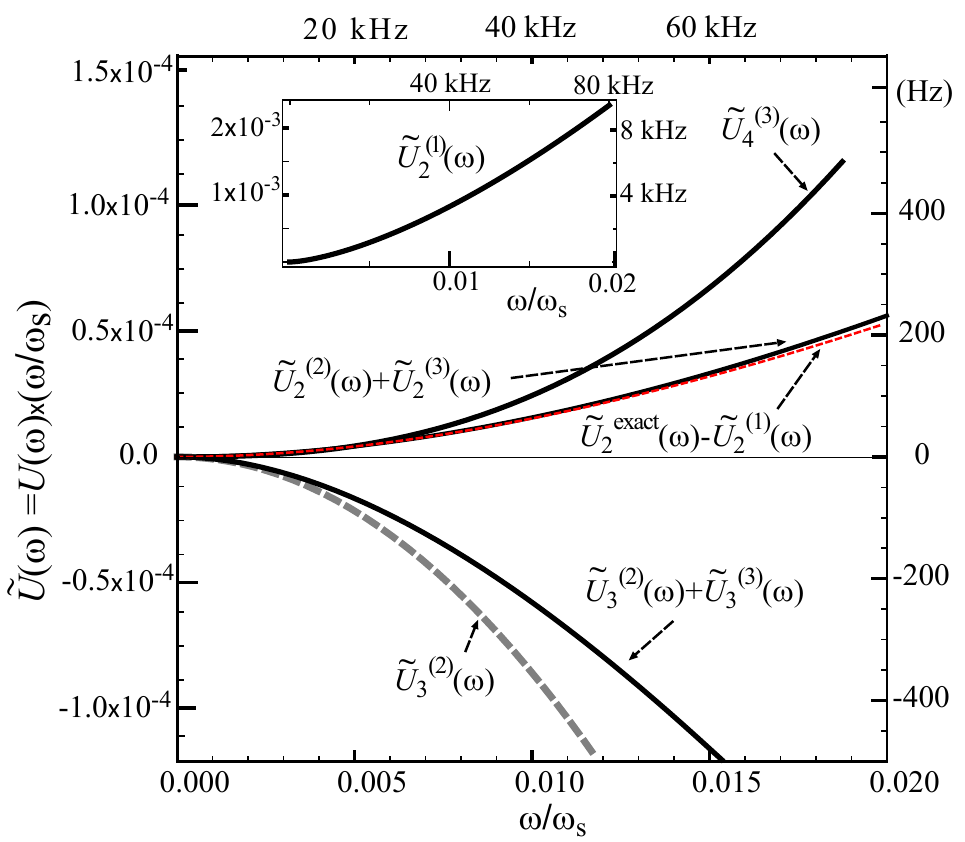} \caption{(Color Online) Perturbative
predictions for dimensionless effective interaction energies $\tilde{U}%
_{m}(\omega)$ versus $\omega/\omega_{\text{s}}$ for fixed scattering length
$a_{\text{t}}(0),$ in units of the energy $\hbar\omega_{\text{s}}=\hbar
^{2}/m_{\text{A}}[a_{\text{t}}(0)]^{2}.$ The inset shows the first-order
two-body energy $\tilde{U}_{2}^{(1)}(\omega).$ The main figure shows the
second- and third-order corrections to the two-, three-, and four-body
energies, assuming no finite-range corrections. The top and right axes in both
figures show the energies converted to frequency units by multiplying by
$\omega_{\text{s}}/2\pi,$ assuming $^{87}$Rb with $a_{\text{t}}(0)=5.3$ nm,
$m_{\text{A}}=86.9$ u, and $\omega_{\text{s}}/2\pi=4.14$ MHz. The line labeled
$\tilde{U}_{2}^{\text{exact}}(\omega)-\tilde{U}_{2}^{(1)}(\omega)$ gives
values using the exact two-body results for $\tilde{U}_{2}(\omega)$ from
\cite{Busch1998}. }%
\label{fig:U2U3U4vsw}%
\end{figure}

Figure$~$\ref{fig:U2U3U4vsw} shows, for the case of a zero-range potential
(i.e. $r_{\text{eff}}=0$), the two-body energies $\tilde{U}_{2}^{(1)}(\omega)$
(in the inset) and $\tilde{U}_{2}^{(2)}(\omega)+\tilde{U}_{2}^{(3)}(\omega)$,
the three-body energies $\tilde{U}_{3}^{(2)}(\omega)$ and $\tilde{U}_{3}%
^{(2)}(\omega)+\tilde{U}_{3}^{(3)}(\omega),$ and the four-body energy
$\tilde{U}_{4}^{(3)}(\omega)$ versus $\omega/\omega_{\text{s}}.$ As expected,
$\tilde{U}_{2}^{(1)}(\omega)$ is the largest contribution. The line labeled
$\tilde{U}_{3}^{(2)}(\omega)$ shows the second-order three-body result found
previously in \cite{Johnson2009}, due to the $c_{3}^{(2)}$ coefficient, and
the line $\tilde{U}_{2}^{(2)}(\omega)+\tilde{U}_{2}^{(3)}(\omega)$ shows the
scale of the correction from the third-order coefficient $c_{3}^{(3)}$. The
effective three- and four-body energies have opposite signs and are of similar
magnitude. Finally, the line labeled $\tilde{U}_{2}^{\text{exact}}%
(\omega)-\tilde{U}_{2}^{(1)}(\omega)$ shows the good agreement with the exact
two-body results from \cite{Busch1998} for the regularized zero-range potential.

It is interesting to directly compare the relative sizes of the second- and
third-order corrections for $^{87}$Rb in a trap. For small magnetic field
strengths, the $^{87}$Rb scattering length and effective range are
approximately $5.3$ nm and $7.9$ nm, respectively \cite{Chin2010}. For a trap
frequency of $10^{2}$ Hz, and thus $a_{\text{t}}(0)/\sigma(\omega)=0.005$
(\textquotedblleft weak\textquotedblright\ confinement), the third-order
two-body terms $c_{2}^{(3)}[a_{\text{t}}(0)/\sigma(\omega)]^{3}$ and
$d_{2}^{(1,2)}[a_{\text{t}}(0)/\sigma(\omega)]^{2}[r_{\text{eff}}%
(0)/\sigma(\omega)]$ are $1\%$ and $2\%$ of the second-order two-body
contribution $c_{2}^{(2)}[a_{\text{t}}(0)/\sigma(\omega)]^{2}.$ Similarly, the
third-order three- and four-body terms $c_{3}^{(3)}[a_{\text{t}}%
(0)/\sigma(\omega)]^{3}$ and $c_{4}^{(3)}[a_{\text{t}}(0)/\sigma(\omega)]^{3}$
are each about $1.5\%$ of the second-order three-body contribution
$c_{3}^{(2)}[a_{\text{t}}(0)/\sigma(\omega)]^{2}.$

For a trap frequency of $10^{4}$ Hz, and thus $a_{\text{t}}(0)/\sigma
(\omega)=0.05$ (\textquotedblleft strong\textquotedblright\ confinement), the
third-order two-body terms increase giving approximately $10\%$ and $20\%$
corrections compared to the second-order two-body contribution. Similarly, the
third-order three- and four-body terms increase giving approximately $15\%$
corrections compared to the second-order three-body contribution. (Notice,
however, that the third-order effective two-body coefficient and the
finite-range coefficient have opposite signs, and hence their contributions
partially cancel.) In typical optical lattice collapse-and-revival experiments
with $^{87}$Rb the confinement is even stronger and the ratio $a_{\text{t}%
}(0)/\sigma(\omega)$ is on the order of $0.05-0.10$
\cite{Will2010,Greiner2002b,Anderlini2006,Sebby2007}. In this regime we expect
non-perturbative effects to also become increasingly important.

\section{Comparison of perturbative energies with energies for finite-range
interactions\label{sec:comparisonofenergies}}

This section compares the predictions of the perturbative ground-state
energies for a zero-range $\delta$-function interaction potential, summarized
in Sec.~\ref{sec:overview} and derived in
Secs.~\ref{sec:1stand2ndordereffectiveinteractions}
and~\ref{sec:thirdorderenergies}, and numerically obtained energies for
$N$-boson systems with finite-range interactions. We show that the
leading-order contribution of an energy-dependent pseudopotential accurately
captures the finite-range effects, and allows us to also validate the analytic
and numerical coefficients found from the zero-range perturbation theory.

We use a finite-range interaction model based on a Gaussian two-body potential
$V_{\text{g}}(r)=V_{0}\exp[-(r/r_{0})^{2}/2]$ with depth (or height) $V_{0}$
and width $r_{0}$ \cite{Stecher2008,Suzuki1998}. For a given width $r_{0}$, we
adjust the depth $V_{0}$ such that $V_{\text{g}}(r)$ produces the physical
free-space $s$-wave scattering length $a_{\text{t}}(0)$ at zero collision
energy. We restrict ourselves to depths $V_{0}$ for which $V_{\text{g}}$
supports no two-body $s$-wave bound state in free-space. This implies that
$V_{0}$ is positive for $a_{\text{t}}(0)>0$ and negative for $a_{\text{t}%
}(0)<0$.

\begin{figure}[ptb]
\includegraphics[width=1\columnwidth]{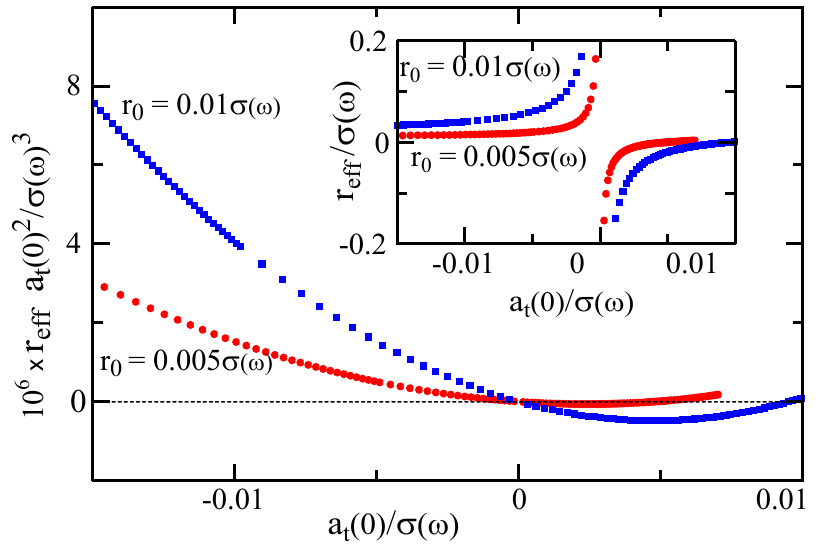}\label{fig:effrangevsat}\caption{(Color
online) Free-space scattering quantities for the Gaussian model potential,
with all lengths expressed in units of $\sigma(\omega).$ Circles and squares
show the volume $r_{\text{eff}}[a_{\text{t}}(0)]^{2}/[\sigma(\omega)]^{3}$ (in
the main figure) and the effective range $r_{\text{eff}}/\sigma(\omega)$ (in
the inset) as a function of $a_{\text{t}}(0)/\sigma(\omega)$ for the Gaussian
potential with $r_{0}=0.005\sigma(\omega)$ and $r_{0}=0.01\sigma(\omega)$,
respectively.}%
\end{figure}

An energy-dependent \emph{free-space} scattering length for two particles with
relative energy $E_{\text{rel}}$ and relative wave number $k_{\text{rel}%
}=\sqrt{m_{\text{A}}E_{\text{rel}}}/\hbar$ can be defined as%
\begin{equation}
a_{\text{f}}(E_{\text{rel}})=-\frac{\tan(\delta_{\text{f}}(k_{\text{rel}}%
))}{k_{\text{rel}}}, \label{phaseshift}%
\end{equation}
where $\delta_{\text{f}}(k_{\text{rel}})$ is the free-space $s$-wave phase
shift. The effect of a finite-range potential on the free-space scattering of
two ultracold bosons can be captured by Taylor-expanding $\delta_{\text{f}%
}(k_{\text{rel}})$ \cite{Taylor2002,mott_theory_1965}, giving%
\begin{equation}
a_{\text{f}}(E_{\text{rel}})=a_{\text{t}}(0)+\frac{1}{2}r_{\text{eff}%
}[a_{\text{t}}(0)]^{2}k_{\text{rel}}^{2}+\cdots,
\label{EnergyDepFreeSpaceScattLength}%
\end{equation}
where $r_{\text{eff}}$ is the effective range parameter which describes the
lowest-order energy dependence of the phase shift \cite{Blume2002,Bolda2002}.

Figure~\ref{fig:effrangevsat} shows the effective range $r_{\text{eff }}$ and
the \textquotedblleft volume\textquotedblright\ $r_{\text{eff}}[a_{\text{t}%
}(0)]^{2}$ for two bosons interacting with the Gaussian potential with two
different choices of $r_{0}/\sigma(\omega)\ll1.$ (The volume factor here
characterizes the leading-order effective-range correction to $s$-wave
scattering.) We extract $r_{\text{eff}}$ by fitting the numerically evaluated
$-\tan(\delta_{\text{f}}(k_{\text{rel}}))/k_{\text{rel}}$ to the
right-hand-side of Eq.~(\ref{EnergyDepFreeSpaceScattLength}) for small
scattering energies. The effective range is positive for negative
$a_{\text{t}}(0),$ negative for small positive $a_{\text{t}}(0),$ and diverges
as $a_{\text{t}}(0)\rightarrow0$. Importantly, since $a_{\text{t}}(0)=0$
implies $V_{0}=0$ (no scattering potential), the volume $r_{\text{eff}%
}[a_{\text{t}}(0)]^{2}\ $also vanishes when $a_{\text{t}}(0)=0$, as seen in
the main part of Fig.~\ref{fig:effrangevsat}. The divergent behavior of the
effective range is also observed for realistic van der Waals
potentials~\cite{gao1998} and indeed for any potential that falls off faster
than $1/r^{5}$ \cite{mott_theory_1965}, although for these potentials (unlike
the Gaussian) $r_{\text{eff}}[a_{\text{t}}(0)]^{2}$ is finite but non-zero in
the limit $a_{\text{t}}(0)\rightarrow0.$

We determine the ground-state energy of $N=3$ and $N=4$ bosons interacting
through the Gaussian model potential under external spherically symmetric
harmonic confinement using a basis set expansion that expresses the relative
$N$-body wave function in terms of explicitly correlated
Gaussians~\cite{Suzuki1998}%
\begin{equation}
\psi_{\text{rel}}=\sum_{k=1}^{N_{b}}u_{k}{\mathcal{S}}\exp\left[  -\frac{1}%
{2}\sum_{i<j}\left(  \frac{r_{ij}}{v_{ij}^{(k)}}\right)  ^{2}\right]  .
\end{equation}
The $u_{k}$ denote expansion coefficients, $N_{b}$ is the number of basis
functions, and ${\mathcal{S}}$ symmetrizes the wave function under the
exchange of any pair of bosons. The $N_{b}\times N(N-1)/2$ variational widths
$v_{ij}^{(k)},$ chosen stochastically from the interval $[r_{0}/5,4\sigma
(\omega)],$ are optimized semi-stochastically following the scheme outlined in
Ref.~\cite{Suzuki1998}. In brief, the variational method works as follows.
Assume we have a basis set consisting of $j-1$ basis functions that yields a
ground-state energy estimate $E_{j-1}$. To add the $j^{th}$ basis function
$(j\leq N_{b}$), we generate a few thousand trial functions. For each trial
function, we solve for a trial ground-state energy by diagonalizing a $j\times
j$ dimensional generalized eigenvalue problem. (It is a generalized eigenvalue
problem because the basis functions are nonorthogonal.) We choose as the
$j^{th}$ basis function the one which makes $E_{j}$ smallest, and repeat this
process for the $(j+1)^{th}$ basis function until $j=N_{b}$. A key benefit of
the explicitly correlated basis functions is that the Hamiltonian and overlap
matrix elements have compact analytical expressions~\cite{Suzuki1998}.

\begin{figure}[ptb]
\includegraphics[width=1\columnwidth]{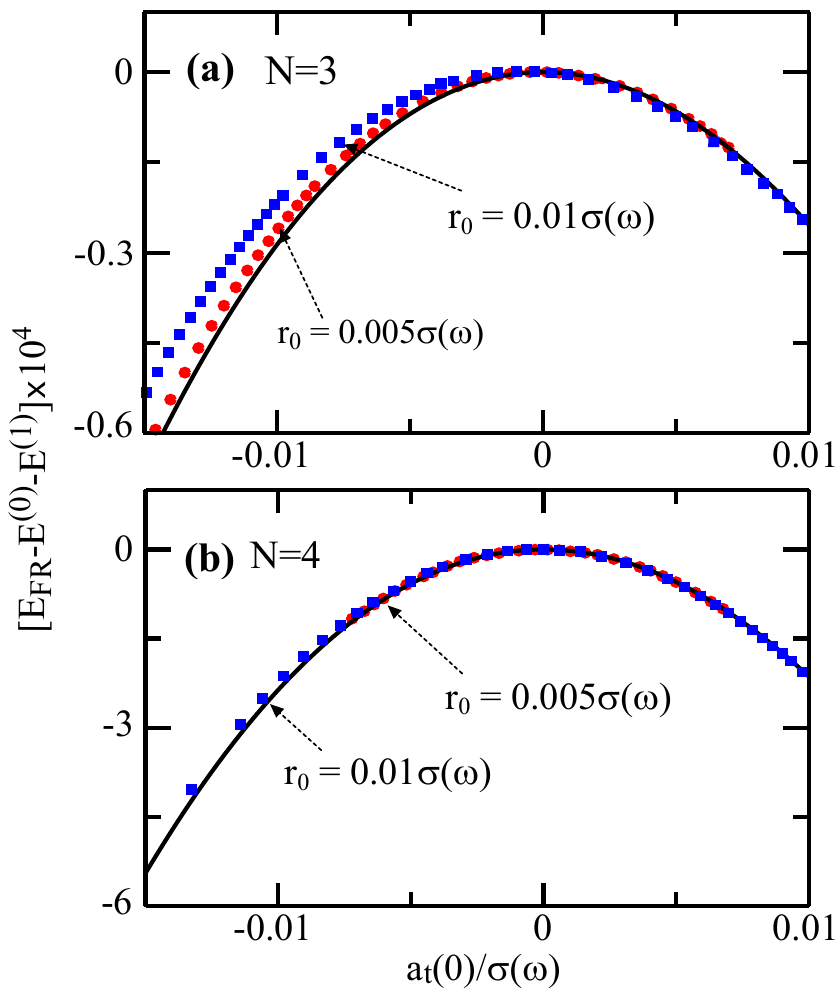} \caption{(Color online) The quantity
$[E_{\text{FR}}-E^{(0)}-E^{(1)}]\times10^{4}$ versus $a_{\text{t}}%
(0)/\sigma(\omega)$ for $N=3$ and $4$ in panel (a) and (b), respectively.
(Energies are in units of $\hbar\omega.$) The finite-range energies
$E_{\text{FR }}$ are numerically computed with $r_{0}=0.01\sigma(\omega)$
(blue squares) and $r_{0}=0.005\sigma(\omega)$ (red circles). The solid line
is $[E^{(2)}+E^{(3)}]\times10^{4}$ found from the perturbative theory with the
zero-range potential.}%
\label{fig:fewbosonenergies1}%
\end{figure}

Convergence is analyzed by investigating the dependence of the energies on
$N_{b}$ and by performing calculations for different sets of widths
$v_{ij}^{(k)}$. To meaningfully compare numerical three- and four-body
energies $E_{\text{FR}}$ for the finite-range (FR) interaction potential with
perturbative results up to order $[a_{\text{t}}(0)]^{3}$, the numerical
accuracy of the finite-range energies should be notably better than
$|a_{\text{t}}(0)/\sigma(\omega)|^{3}$. For example, for $|a_{\text{t}%
}(0)|=0.001\sigma(\omega)$ and $|a_{\text{t}}(0)|=0.01\sigma(\omega),$ this
implies numerical accuracy better than $10^{-9}$ and $10^{-6}$, respectively.
An analysis of the basis set error shows that our $N$-body energies are
sufficiently accurate to test the perturbative predictions up to order
$[a_{\text{t}}(0)]^{3}$ for $|a_{\text{t}}(0)|\gtrsim0.1r_{0},$ using about
$100$ and $500$ basis functions for $N=3$ and $N=4$, respectively. Our
numerical accuracy is insufficient to test the perturbative predictions for
smaller $|a_{\text{t}}(0)|$.

\begin{figure}[ptb]
\includegraphics[width=1\columnwidth]{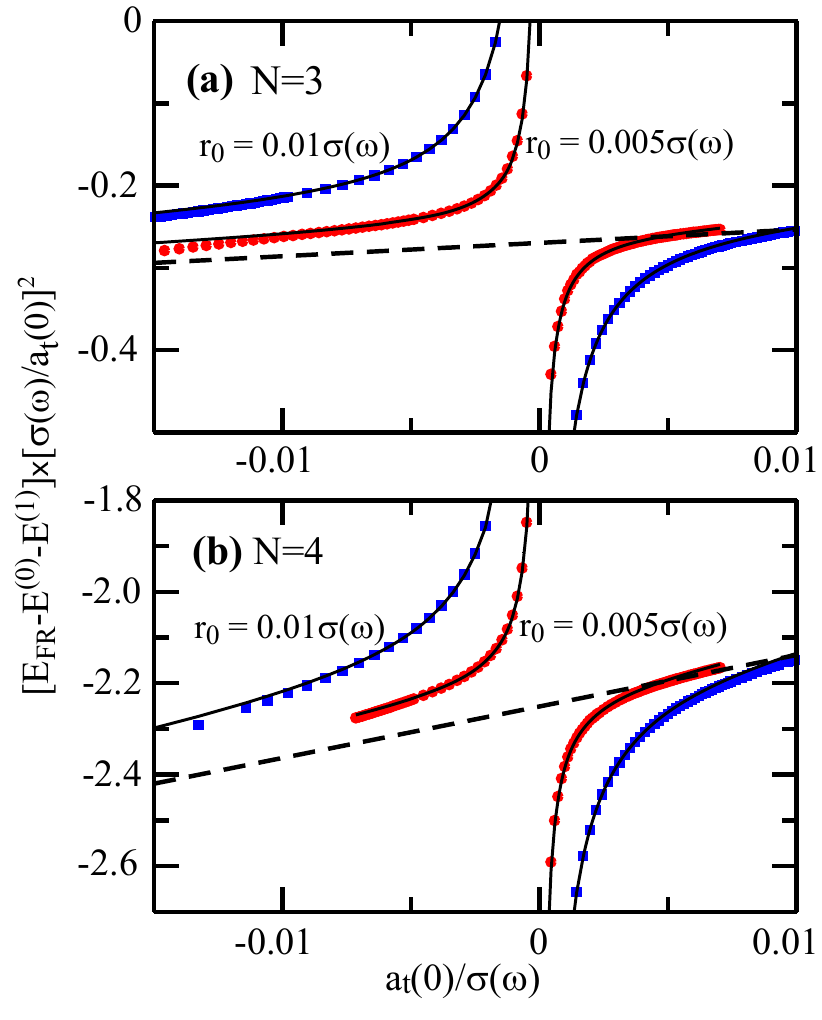}
\caption{(Color online) Analysis of the three-boson [panel (a)] and four-boson
[panel (b)] energies including scattering length and effective range effects.
All energies are scaled by $[\sigma(\omega)/a_{\text{t}}(0)]^{2}$ to emphasize
the corrections due to finite range effects. The numerically determined
finite-range energies $E_{\text{FR}}$ are calculated for the Gaussian
potential with spatial width $r_{0}=0.005\sigma(\omega)$ (red circles) and
$r_{0}=0.010\sigma(\omega)$ (blue squares), respectively. The dashed line
shows the scaled perturbation theory prediction $E^{(2)}+E^{(3)}$ for a
zero-range, delta-function potential. The divergence at $a_{\text{t}%
}(0)/\sigma(\omega)=0$ is due to the divergence of the effective range at zero
scattering length. The unscaled energy shift vanishes when $a_{\text{t}%
}(0)=0.$ The solid lines show the scaled energies $E^{(2)}+E^{(3)}+E^{(1,2)},
$ which include the perturbatively calculated finite-range correction
$E^{(1,2)}$.}%
\label{fig:fewbosonenergies2}%
\end{figure}

Figure~\ref{fig:fewbosonenergies1} shows the quantity $[E_{\text{FR}}%
-E^{(0)}-E^{(1)}]\times10^{4}$ versus $a_{\text{t}}(0)/\sigma(\omega),$ with
the finite-range energies $E_{\text{FR }}$ numerically computed using
$r_{0}=0.01\sigma(\omega)$ (the blue squares) and $r_{0}=0.005\sigma(\omega)$
(the red circles). We have subtracted the energies $E^{(0)}$ and $E^{(1)}$
obtained from the perturbative theory to better examine the physics beyond
first order in $a_{\text{t}}(0).$ The solid line is $[E^{(2)}+E^{(3)}%
]\times10^{4}$ from the perturbative theory with $r_{\text{eff}}=0$. Panels
(a) and (b) give the energies for $N=3$ and $4$ bosons, respectively. For
$N=3,$ we see that finite-range corrections to the zero-range theory become
more significant for increasing $r_{0}.$

In Figs.~\ref{fig:fewbosonenergies2}(a) and (b), we multiply the $N=3$ and $4
$ energies $E_{\text{FR}}-E^{(0)}-E^{(1)}$ by $[\sigma(\omega)/a_{\text{t}%
}(0)]^{2}$. The perturbative predictions for $(E^{(2)}+E^{(3)})[\sigma
(\omega)/a_{\text{t}}(0)]^{2}$ are straight lines. The nonperturbative
numerical results are for potentials with $r_{0}=0.005\sigma(\omega)$ and
$r_{0}=0.01\sigma(\omega)$. The figures show that the scaled numerical results
are singular near zero scattering length, and only approach the zero-range
perturbative results with increasing $|a_{\text{t}}(0)|.$ Moreover, by
decreasing $r_{0}$ the difference between the perturbative results and the
scaled finite-range energies is reduced, and we conclude that the divergences
at $a_{\text{t}}(0)=0$ are due to the finite range of the Gaussian potential.
Multiplying the energies by $[\sigma(\omega)/a_{\text{t}}(0)]^{2}$ has
magnified the finite-range corrections, showing that an effective field theory
description for finite-range potentials requires corrections to the zero-range
$\delta$-function potential.

We can calculate the leading-order influence of a finite-range potential by
including the \emph{energy-dependent} zero-range pseudopotential of
Eq.~(\ref{V2prime}). For the $N$-boson ground state, the pseudopotential gives
to first order in $g_{2}^{\prime}$ an energy shift $E^{(1,2)}$ [see
Eq.~(\ref{E12})]. At this order, the addition of $\mathcal{V}_{2}^{\prime}$ is
equivalent to replacing $a_{\text{t}}(0)\ $by $a_{\text{f}}(E_{\text{rel}})$,
with Eq.~(\ref{EnergyDepFreeSpaceScattLength}) evaluated at the relative
zero-point energy $E_{\text{rel}}=3/2$ (in units of $\hbar\omega$) of two
non-interacting bosons in the trap.

The solid lines in Fig.~\ref{fig:fewbosonenergies2} show $(E^{(2)}%
+E^{(3)}+E^{(1,2)})[\sigma(\omega)/a_{\text{t}}(0)]^{2}$ as a function of
$a_{\text{t}}(0)/\sigma(\omega)$ for $N=3$ and $N=4$ trapped bosons,
respectively. Combining the perturbative predictions for zero-range
contributions $E^{(2)}+E^{(3)}$ and the effective-range correction $E^{(1,2)}$
gives excellent agreement with the nonperturbative finite-range energies. The
comparison validates the perturbation theory and predictions derived in this
paper for effective interactions including finite-range corrections, through
third order in $a_{\text{t}}(0)$. It also shows that the divergences in
Fig.~\ref{fig:fewbosonenergies2} at $a_{\text{t}}(0)=0$ are due to the
divergence of the effective range shown in the inset of
Fig.~\ref{fig:effrangevsat}. Finally, we note that the energy shift is
proportional to the volume $r_{\text{eff}}[a_{\text{t}}(0)]^{2}$ and goes to
zero at $a_{\text{t}}(0)=0,$ as expected.

\section{First- and second-order effective
interactions\label{sec:1stand2ndordereffectiveinteractions}}

\subsection{Hamiltonian and renormalization condition}

The numerical results in Sec.~\ref{sec:comparisonofenergies} show that
finite-range effects are important at third order in perturbation theory for
realistic bosons. We incorporate these corrections by modeling the pairwise
collisions of ultracold bosons by combining the zero-range pseudopotential%
\begin{equation}
\mathcal{V}_{2}(\mathbf{r}_{1}-\mathbf{r}_{2})=4\pi\frac{\hbar^{2}%
}{m_{\text{A}}}a_{\text{bare}}\delta^{(3)}(\mathbf{r}_{1}-\mathbf{r}%
_{2}),\label{twobodypseudopotential}%
\end{equation}
where $a_{\text{bare}}$ is now identified as the bare scattering length, and
the effective-range potential

\begin{multline}
\mathcal{V}_{2}^{\prime}(\mathbf{r}_{1}-\mathbf{r}_{2})=-\frac{1}%
{2}g_{2,\text{bare}}^{\prime} \\ \times [\overleftarrow{\nabla}_{12}^{2}\delta
^{(3)}(\mathbf{r}_{1}-\mathbf{r}_{2})+\delta^{(3)}(\mathbf{r}_{1}%
-\mathbf{r}_{2})\overrightarrow{\nabla}_{12}^{2}],
\end{multline}
which has the bare coupling constant
\begin{equation}
g_{2,\text{bare}}^{\prime}=\left(  4\pi\frac{\hbar^{2}}{m_{\text{A}}}\right)
\left(  \frac{1}{2}r_{\text{eff}}[a_{\text{bare}}]^{2}\right)  .
\end{equation}
The interactions of $N$ ultracold neutral bosons can be described in quantum
field theory with the Hamiltonian $\mathcal{H}=\mathcal{H}_{0}+\mathcal{H}%
_{\text{I}},$ where $\mathcal{H}_{0}$ is the single-particle Hamiltonian and
\begin{multline}
\mathcal{H}_{\text{I}}=\frac{1}{2}\int\hat{\psi}^{\dag}(\mathbf{r}_{1}%
)\hat{\psi}^{\dag}(\mathbf{r}_{2})[\mathcal{V}_{2}(\mathbf{r}_{1}%
-\mathbf{r}_{2})\\
+\mathcal{V}_{2}^{\prime}(\mathbf{r}_{1}-\mathbf{r}_{2})]\hat{\psi}%
(\mathbf{r}_{1})\hat{\psi}(\mathbf{r}_{2})d\mathbf{r}_{1}d\mathbf{r}%
_{2}.
\end{multline}
The field operators $\hat{\psi}(\mathbf{r)}$ and $\hat{\psi}^{\dag
}(\mathbf{r)}$ respectively annihilate and create a boson at position
$\mathbf{r}$. We assume the absence of intrinsic three- or higher-body interactions.

The bosonic field is expanded over isotropic harmonic oscillator states with
frequency $\omega$ as
\begin{equation}
\hat{\psi}(\mathbf{r)}=\sum_{nlm}\phi_{nlm}(\mathbf{r)}\hat{a}_{nlm}=\sum
_{i}\phi_{i}(\mathbf{r)}\hat{a}_{i}, \label{psi}%
\end{equation}
with $\hat{a}_{i}$ annihilating a boson in orbital $\phi_{i}(\mathbf{r)}.$ In
the following we use the shorthand notation $i=\{nlm\}$, denoting the
(dimensionless) single-particle energies as $\varepsilon_{i}=\varepsilon
_{nlm}=(2n+l+3/2),$ where $n,l=0,1,2,...,$ and $i=0$ is the $\{nlm\}=\{000\}$
single-particle vibrational ground state. Substituting Eq.~(\ref{psi}) into
$\mathcal{H}$ and dividing by $\hbar\omega,$ we define the dimensionless
Hamiltonian $H=H_{0}+H_{\text{I}}+H_{\text{I}}^{\prime},$ where $H_{0}%
=\sum_{i}\varepsilon_{i}\hat{a}_{i}^{\dag}\hat{a}_{i},$
\begin{equation}
H_{\text{I}}=\frac{1}{2}\left(  \frac{a_{\text{bare}}}{\sigma(\omega)}\right)
\sum_{ijkl}K_{ij;kl}\hat{a}_{i}^{\dag}\hat{a}_{j}^{\dag}\hat{a}_{k}\hat{a}%
_{l}, \label{Hintbare}%
\end{equation}
and%
\begin{equation}
H_{\text{I}}^{\prime}=\frac{1}{2}\left(  \frac{1}{2}\frac{r_{\text{eff}%
}[a_{\text{bare}}]^{2}}{\sigma(\omega)^{3}}\right)  \sum_{ijkl}K_{ij;kl}%
^{\prime}\hat{a}_{i}^{\dag}\hat{a}_{j}^{\dag}\hat{a}_{k}\hat{a}_{l}.
\label{Hintbareprime}%
\end{equation}
The matrix elements
\begin{equation}
K_{ij;kl}=4\pi\lbrack\sigma(\omega)]^{3}\int\phi_{i}^{\ast}(\mathbf{r)}%
\phi_{j}^{\ast}(\mathbf{r)}\phi_{k}(\mathbf{r)}\phi_{l}(\mathbf{r)}d\mathbf{r}
\label{matrix elements}%
\end{equation}
and%
\begin{equation}
K_{ij;kl}^{\prime}=-4\pi\lbrack\sigma(\omega)]^{5}\int[\phi_{i}^{\ast
}(\mathbf{r})\phi_{j}^{\ast}(\mathbf{r})]\overleftrightarrow{\nabla
}_{\mathbf{r}}^{2}[\phi_{k}(\mathbf{r})\phi_{l}(\mathbf{r})]d\mathbf{r}%
\end{equation}
are normalized such that $K_{00;00}=\sqrt{2/\pi}$ and $K_{00;00}^{\prime
}=(3/4)\sqrt{2/\pi},$ with the semi-colon separating initial and final states
and $\overleftrightarrow{\nabla}_{\mathbf{r}}^{2}=(\overleftarrow{\nabla
}_{\mathbf{r}}^{2}+\overrightarrow{\nabla}_{\mathbf{r}}^{2})/2.$ The factors
of $[\sigma(\omega)]^{3}$ and $[\sigma(\omega)]^{5}$ make the matrix elements
dimensionless and $\omega$-\emph{independent}. As explained in
Sec.~\ref{sec:overview}, we assume a regime where $H_{\text{I}}^{\prime}$ can
be treated as third order in perturbation theory.

The noninteracting ground state containing $N$ bosons in the $i=0$ (i.e.,
$nlm=000$) vibrational ground state is $|N\rangle=\hat{a}_{0}^{\dag
N}|0\rangle/\sqrt{N!}$, with energy $E^{(0)}=N\varepsilon_{0}$ and
$\varepsilon_{0}=3/2.$ First-order perturbation theory in $H_{\text{I}}$ gives
$E^{(1)}(\omega)=(1/2)N(N-1)U_{2}^{(1)}$ with%
\begin{equation}
U_{2}^{(1)}=\alpha_{2}^{(1)}\left(  \frac{a_{\text{bare}}}{\sigma(\omega
)}\right)  , \label{U2bare}%
\end{equation}
using $\left\langle N\right\vert a_{0}^{\dag}a_{0}^{\dag}a_{0}a_{0}\left\vert
N\right\rangle =N(N-1)$ and recalling that $|N\rangle$ denotes $N$ bosons in
the non-interacting vibrational state $nlm=000.$ The two-body, first-order
coefficient is $\alpha_{2}^{(1)}=\sqrt{2/\pi}.$

At higher orders in $H_{\text{I}},$ there are divergences due to the $\delta
$-function potential (see e.g.~\cite{Huang1957,Fetter1971}). We regulate these
by either truncating sums over intermediate states at a high-energy cutoff
$\hbar\omega_{c},$ or by using an exponential regulator function. The former
is more convenient for numerical approximations, while the latter is more
convenient for analytic results. In either case, we find at second order that
$U_{2}^{(2)}$ diverges as $\sqrt{\omega_{c}}$ and renormalization is required.
Although this can be done using \emph{bare }perturbation theory, in which
infinities are absorbed by appropriately redefining bare parameters, we use
the method of renormalized perturbation theory which provides a systematic and
self-consistent approach for calculations beyond second order involving
multiple divergent terms.

Renormalized perturbation theory (e.g., see~\cite{Srednicki2007}) re-expresses
the bare scattering length as%
\begin{equation}
a_{\text{bare}}=a_{\text{t}}(\omega_{0})+a_{\text{ct}}(\omega_{0}).
\label{abare}%
\end{equation}
A \emph{renormalization condition }defines $a_{\text{t}}(\omega_{0})$ as the
physical scattering length for two bosons in a trap at frequency $\omega_{0}$.
The cutoff dependent remainder $a_{\text{ct}}(\omega_{0})$ is called a
counterterm. For brevity, this notation suppresses the dependence of
$a_{\text{ct}}(\omega_{0})$ on $\omega_{c}$. In the following, we call
$a_{\text{t}}(\omega_{0})$ the \textquotedblleft trap scattering
length\textquotedblright\ at frequency $\omega_{0},$ to distinguish it from
the energy-dependent \emph{free-space} scattering length $a_{\text{f}%
}(E_{\text{rel}}$) defined in Eq.~(\ref{phaseshift}). In the limits of zero
relative collision energy and $\omega_{0}=0,$ the \emph{trap} and
\emph{free-space} scattering lengths are equal, i.e., $a_{\text{t}%
}(0)=a_{\text{f}}(0).$ With the combination of $\mathcal{V}_{2}$ and
$\mathcal{V}_{2}^{\prime},$ the trap scattering length $a_{\text{t}}%
(\omega_{0})$ includes both the effects of the $\omega_{0}$-dependent dressing
by quantum fluctuations to higher orbitals and finite-range effects. Note that
the trap scattering length $a_{\text{t}}(\omega_{0})\ $does not, in general,
equal the free-space scattering length $a_{\text{f}}(E_{\text{rel}})$ defined
in Eq.~(\ref{EnergyDepFreeSpaceScattLength}) because the latter does not
correctly capture the influence of the harmonic confinement on the quantum
fluctuations to higher orbitals.

\begin{figure}[ptb]
\includegraphics[width=1\columnwidth]{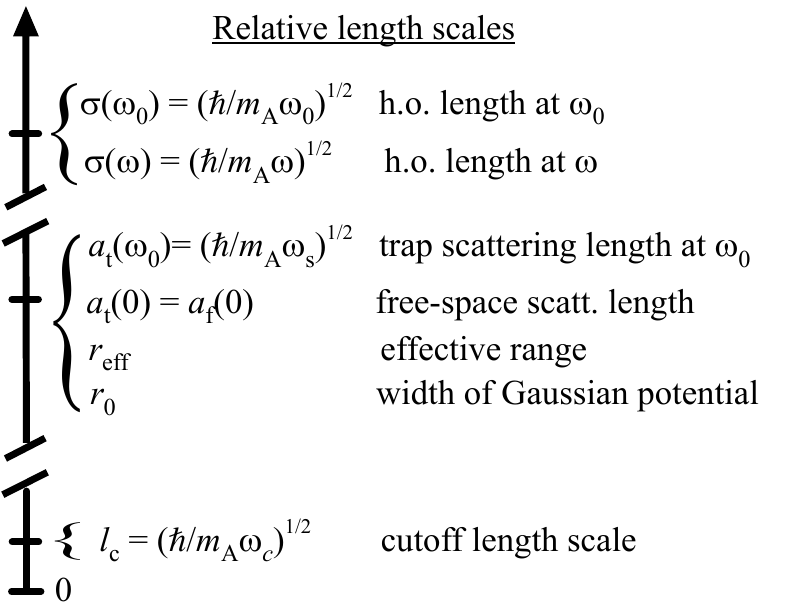} \caption{The length scales of
interacting, harmonically trapped, ultracold bosons. We assume a separation of
length scales $l_{c}\ll a_{\text{t}}(\omega_{0})\ll\sigma(\omega_{0}),$ or
equivalently a separation of energy scales $\omega_{c}\gg\omega_{\text{s}}%
\gg\omega_{0},$ where $\omega_{\text{s}}=\hbar/m_{\text{A}}a_{\text{t}}%
(\omega_{0})^{2}.$ The harmonic oscillator lengths $\sigma(\omega)$ and
$\sigma(\omega_{0})$ are of the same order, although $\sigma(\omega_{0})$ is
not necessarily larger than $\sigma(\omega)$. Similarly, we assume that
$a_{\text{t}}(\omega_{0}),$ $a_{\text{t}}(0),$ $r_{\text{eff}}$, and the
Gaussian width $r_{0}$ are of the same order. The order of length scales
within a group is arbitrary.}%
\label{fig:lengths}%
\end{figure}

Together with the renormalization condition, the other key ingredient in
renormalized perturbation theory is that the leading-order scattering length
counterterm $a_{\text{ct}}(\omega_{0})$ is proportional to $[a_{\text{t}%
}(\omega_{0})]^{2};$ in other words, it is a second- and higher-order
contribution. This, plus the renormalization condition, systematically
reorganizes the perturbation theory, order-by-order, so that it is an
expansion in the physical value $a_{\text{t}}(\omega_{0})$ instead of
$a_{\text{bare}}.$ Figure~\ref{fig:lengths} summarizes the relationship
between the characteristic length and energy scales for our model system of
trapped ultracold bosons.

Substituting Eq.$~$(\ref{abare}) into Eqs.~(\ref{Hintbare}) and
(\ref{Hintbareprime}) gives
\begin{equation}
H_{\text{I}}(\omega;\omega_{0})=V(\omega;\omega_{0})+V^{\prime}(\omega
;\omega_{0})+V_{\text{ct}}(\omega;\omega_{0}), \label{Hint}%
\end{equation}
where the zero-range and counterterm operators are%
\begin{flalign}
& V(\omega;\omega_{0})    =\frac{1}{2}\left(  \frac{a_{\text{t}}(\omega_{0}%
)}{\sigma(\omega)}\right)  \sum_{ijkl}K_{ij;kl}\hat{a}_{i}^{\dag}\hat{a}%
_{j}^{\dag}\hat{a}_{k}\hat{a}_{l},\label{Vomega} & \\
 & V_{\text{ct}}(\omega;\omega_{0})   =\frac{1}{2}\left(  \frac{a_{\text{ct}%
}(\omega_{0})}{\sigma(\omega)}\right)  \sum_{ijkl}K_{ij;kl}\hat{a}_{i}^{\dag
}\hat{a}_{j}^{\dag}\hat{a}_{k}\hat{a}_{l}, \label{Womega}&%
\end{flalign}
and the effective-range operator is%
\begin{flalign}
 V^{\prime}(\omega;\omega_{0}) & =\frac{1}{2}\left(  \frac{1}{2}\frac
{r_{\text{eff}}[a_{\text{t}}(\omega_{0})]^{2}}{[\sigma(\omega)]^{3}}\right)
\sum_{ijkl}K_{ij;kl}^{\prime}\hat{a}_{i}^{\dag}\hat{a}_{j}^{\dag}\hat{a}%
_{k}\hat{a}_{l} & \nonumber \\ &+\mathcal{O}\left(  \frac{r_{\text{eff}}[a_{\text{t}}(0)]^{3}%
}{[\sigma(\omega)]^{4}}\right)  . & \label{Vomegaprime}%
\end{flalign}
The renormalized perturbation theory is then organized based on the
observation that $V(\omega;\omega_{0})$ is proportional to $a_{\text{t}%
}(\omega_{0})$, $V_{\text{ct}}(\omega;\omega_{0})$ is proportional to
$[a_{\text{t}}(\omega_{0})]^{2}$, and $V^{\prime}(\omega;\omega_{0})$ is (for
the regime considered here) proportional to $[a_{\text{t}}(\omega_{0})]^{3}.$
The single counterterm operator $V_{\text{ct}}(\omega;\omega_{0})$ cancels all
divergences from the operator $V(\omega;\omega_{0}),$ at all orders in
perturbation theory. In contrast, the effective-range operator $V^{\prime
}(\omega;\omega_{0})$ leads to a nonrenormalizable field theory with the
consequence that new counterterm operators are required at every order in
perturbation theory beyond first order in $g_{2}^{\prime}$; because we are
only working to first order in $g_{2}^{\prime}$ in this paper, no additional
counterterms are needed.

Note that the frequency $\omega_{0}$ at which $a_{\text{t}}(\omega_{0})$ is
defined and the trap frequency $\omega$ for which we want to compute energies
are independent. In the overview, we summarized our results for the special
case where $\omega_{0}=0$. The general case of arbitrary $\omega_{0}$
facilitates renormalization of the perturbation theory. More importantly, the
renormalized perturbation theory is \textquotedblleft
calibrated\textquotedblright\ to a measured value of $a_{\text{t}}(\omega
_{0})$ at a desired trap frequency $\omega_{0}$, and is then used to predict
energies for trap frequencies $\omega$ not generally equal to $\omega_{0}.$

We can now compute the ground-state energy%
\begin{align}
E(\omega;\omega_{0})  &  =\varepsilon_{0}N+\frac{1}{2!}U_{2}(\omega;\omega
_{0})N(N-1) \label{Eww0} \\ &+ \frac{1}{3!}U_{3}(\omega;\omega_{0})N(N-1)(N-2) \nonumber \\
&  +\frac{1}{4!}U_{4}(\omega;\omega_{0})N(N-1)(N-2)(N-3)+....\nonumber
\end{align}
We have used the semi-colon notation in Eqs.~\eqref{Hint}, \eqref{Vomega}, \eqref{Womega}, \eqref{Vomegaprime}, and \eqref{Eww0} to distinguish between the roles of the frequencies $\omega$ and $\omega_{0}$. Before renormalization, the
interaction energies $U_{m}(\omega;\omega_{0})$, found from perturbation
theory in $H_{\text{I}}(\omega;\omega_{0}),$ are functions of $a_{\text{t}%
}(\omega_{0})$ and $a_{\text{ct}}(\omega_{0}).$ The renormalization condition
can be expressed as
\begin{equation}
U_{2}(\omega=\omega_{0};\omega_{0})=\sqrt{\frac{2}{\pi}}\left(  \frac
{a_{\text{t}}(\omega_{0})}{\sigma(\omega=\omega_{0})}\right)  ,
\label{Renormalization condition}%
\end{equation}
which, in practice, is solved for $a_{\text{ct}}(\omega_{0})$ to the desired
order in perturbation theory. Another way of describing the renormalization
condition is that $a_{\text{ct}}(\omega_{0})$ is tuned such that the
first-order result is exact and the second- and higher-order corrections to
the two-body energy vanish when evaluated for two bosons in a trap with
$\omega=\omega_{0}.$ After renormalization, the interaction energies
$U_{m}(\omega;\omega_{0})$ depend only on $a_{\text{t}}(\omega_{0})$ and,
moreover, the $\omega$-dependence of the ground-state energy satisfies
$E(\omega;\omega_{0})=E(\omega;\omega_{0}^{\prime}),$ for any pair of
frequencies $\omega_{0}$ and $\omega_{0}^{\prime}.$

\subsection{Energy at first-order in scattering length
\label{sec:firstorderenergies}}

We use renormalized Rayleigh-Schr\"{o}dinger (RS) perturbation theory to
compute the $N$-boson ground-state energy $E=\sum_{n=0}E^{(n)},$ where
$E^{(n)}$ is proportional to $[a_{\text{t}}(\omega_{0})]^{n}$. We separate the
contributions at each order into $m$-body energies, such that $U_{m}%
(\omega;\omega_{0})=\sum_{n}U_{m}^{(n)}(\omega;\omega_{0})$. The zeroth-order
term is $E^{(0)}(\omega)=\varepsilon_{0}N$. The first-order energy shift is
\begin{align}
E^{(1)}(\omega;\omega_{0}) &=\langle N|H_{\text{I}}(\omega;\omega_{0}%
)|N\rangle=\langle N|V(\omega;\omega_{0})|N\rangle \nonumber \\ &=\frac{1}{2}\sqrt{\frac
{2}{\pi}}\left(  \frac{a_{\text{t}}(\omega_{0})}{\sigma(\omega)}\right)
N(N-1), \label{E1}%
\end{align}
using the fact that $V$, $V_{\text{ct}},$ and $V^{\prime}$ are $\mathcal{O}%
(a_{\text{t}}(\omega_{0})/\sigma(\omega))$, $\mathcal{O}([a_{\text{t}}%
(\omega_{0})/\sigma(\omega)]^{2})$, and $\mathcal{O(}[a_{\text{t}}(\omega
_{0})/\sigma(\omega)]^{3}),$ respectively, and $\left\langle N\right\vert
a_{0}^{\dag}a_{0}^{\dag}a_{0}a_{0}\left\vert N\right\rangle =N(N-1)$.

Comparing to Eq.~(\ref{Eww0}), we see that the two-body energy to first-order
for \emph{any} $\omega$ and $\omega_{0}$ is
\begin{equation}
U_{2}^{(1)}(\omega;\omega_{0})=c_{2}^{(1)}\left(  \frac{a_{\text{t}}%
(\omega_{0})}{\sigma(\omega)}\right)  ,
\end{equation}
with%
\begin{equation}
c_{2}^{(1)}=\alpha_{2}^{(1)}=\sqrt{\frac{2}{\pi}}.
\end{equation}
For a trap with $\omega=\omega_{0},$ the renormalization condition says that
$U_{2}^{(1)}(\omega_{0};\omega_{0})=\sqrt{2/\pi}[a_{\text{t}}(\omega
_{0})/\sigma(\omega_{0})]$ is the exact two-body energy. For $\omega\neq
\omega_{0},$ $U_{2}^{(1)}(\omega\neq\omega_{0};\omega_{0})$ is the leading
order contribution to the full two-body energy $U_{2}(\omega;\omega_{0}),$
but, as shown in the following sections, there are higher-order corrections
that become increasingly important the more $\omega$ differs from $\omega
_{0}.$

\subsection{Energy at second-order in scattering length
\label{sec:secondorderenergies}}

The second-order energy shift is given by
\begin{equation}
E^{(2)}=V_{\text{ct;}00,00}-\sum_{ij\neq00}^{\omega_{c}/\omega}\frac
{V_{00,ij}V_{ij,00}}{\Delta\varepsilon_{ij}}, \label{2ndorderPTformula}%
\end{equation}
where $V_{ij,kl}=\langle ij|V|kl\rangle$ and $V_{\text{ct;}ij,kl}=\langle
ij|V_{\text{ct}}|kl\rangle.$ The notation $|ij\rangle=Z_{ij}\hat{a}_{i}^{\dag
}\hat{a}_{j}^{\dag}\hat{a}_{0}\hat{a}_{0}\left\vert N\right\rangle $ denotes
the state with either one or two particles excited from the non-interacting
ground state, $\Delta\varepsilon_{ij}=\varepsilon_{i}+\varepsilon
_{j}-2\varepsilon_{0},$ $Z_{ij}$ is a normalization factor, and $ij\neq00$
denotes summing over all $i,j$ except $i=j=0$.
Equation~(\ref{2ndorderPTformula}) is modified from the usual RS perturbation
theory because of the presence of the $\mathcal{O}([a_{\text{t}}(\omega
_{0})]^{2})$ interaction term $V_{\text{ct}},$ which generates the counterterm contribution.

The sums over intermediate states $\left\vert ij\right\rangle $ exclude the
ground state $i=j=0$, and are regularized using either a hard cutoff
$\Delta\varepsilon_{ij}<\omega_{c}/\omega,$ or an exponential regulator
$\Delta\varepsilon_{ij}^{-1}\rightarrow e^{-\Delta\varepsilon_{ij}%
\omega/\omega_{c}}\Delta\varepsilon_{ij}^{-1},$ where $\hbar\omega_{c}$ is a
high-energy cutoff. In the limit $\omega_{c}/\omega\rightarrow\infty,$ these
regulators are equivalent.

Using Eqs. (\ref{Vomega}) and (\ref{Womega}), we have
\begin{align}
& E^{(2)}(\omega;\omega_{0}) = \frac{1}{2}\alpha_{2}^{(1)}\left(  \frac
{a_{\text{ct}}(\omega_{0})}{\sigma(\omega)}\right)  N(N-1) \label{2ndOrderEnergy} \\  &-\frac{N(N-1)}{4}\left(
\frac{a_{\text{t}}(\omega_{0})}{\sigma(\omega)}\right)^{2} \sum
_{ij\neq00,kl}^{\omega_{c}/\omega}\frac{K_{00;ij}K_{kl;00}}{\Delta
\varepsilon_{ij}}\langle\hat{a}_{i}\hat{a}_{j}\hat{a}_{k}^{\dagger}\hat{a}%
_{l}^{\dagger}\rangle, \nonumber 
\end{align}
and the expectation value is with respect to the non-interacting ground state
$|N-2\rangle\propto\hat{a}_{0}\hat{a}_{0}|N\rangle$. The notation
$ij\neq00,kl$ indicates that the sum is over all $i,j,k,l$ except $i=j=0.$
Wick's theorem gives
\begin{align}
\langle\hat{a}_{i}\hat{a}_{j}\hat{a}_{k}^{\dagger}\hat{a}_{l}^{\dagger}\rangle
&  =4\langle\text{:}\contraction{}{\hat{a}}{_{a}\hat{a}_{b}}{\hat{a}}\hat
{a}_{i}\hat{a}_{j}\hat{a}_{k}^{\dag}\hat{a}_{l}^{\dag}\text{:}\rangle
+2\langle\text{:}%
\contraction{}{\hat{a}}{_{a}\hat{a}_{b}}{\hat{a}}\contraction[2ex]{\hat{a}_{a}}{\hat{a}}{_{b}\hat{a}_{c}^{\dag}}{\hat{a}}\hat
{a}_{i}\hat{a}_{j}\hat{a}_{k}^{\dag}\hat{a}_{l}^{\dag}\text{:}\rangle
\nonumber\\
&  =4\delta_{ik}(N-2)+2\delta_{ik}\delta_{jl}, \label{wick2ndorder}%
\end{align}
where :: denotes normal ordering, uncontracted indices are set to zero, and
contractions $\contraction{}{\hat{a}}{_{a}}{\hat{a}}\hat{a}_{i}\hat{a}%
_{k}^{\dag}=\delta_{ik}$. Also, we have used $\langle$:$\hat{a}_{i}\hat{a}%
_{j}\hat{a}_{k}^{\dag}\hat{a}_{l}^{\dag}$:$\rangle=0$, $\langle$:$\hat{a}%
_{0}^{\dag M}\hat{a}_{0}^{M}$:$\rangle=(N-2)(N-3)...(N-M+1),$ and combined
equivalent terms. Because of the factor $N-2,$ the first term of
Eq.~(\ref{wick2ndorder}) can be understood as leading to an effective
three-body interaction, whereas the second term is a correction to the
two-body interaction.

The second-order interaction energies $U_{2}^{(2)}(\omega;\omega_{0})$ and
$U_{3}^{(2)}(\omega;\omega_{0})$ can be extracted by evaluating
Eq.~(\ref{2ndOrderEnergy}) and comparing with Eq.~(\ref{Eww0}). This gives
\begin{align}
U_{2}^{(2)}(\omega;\omega_{0}) & =\alpha_{2}^{(1)}\left(  \frac{a_{\text{ct}%
}(\omega_{0})}{\sigma(\omega)}\right)  -\beta_{2}^{(2)}(\omega)\left(
\frac{a_{\text{t}}(\omega_{0})}{\sigma(\omega)}\right)  ^{2}%
\nonumber \\ &=\raisebox{-0.05in}{\includegraphics[]{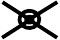}}-\raisebox{-0.05in}{\includegraphics[]{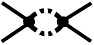}},
\label{U22}%
\end{align}
and%
\begin{equation}
U_{3}^{(2)}(\omega;\omega_{0})=-6\alpha_{3}^{(2)}\left(  \frac{a_{\text{t}%
}(\omega_{0})}{\sigma(\omega)}\right)  ^{2}%
=-6\raisebox{-0.05in}{\includegraphics[]{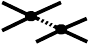}}. \label{U32}%
\end{equation}

The expressions for $\beta_{2}^{(2)}(\omega)$ and $\alpha_{3}^{(2)}$ are
defined in Table~\ref{table:alphaandbeta}, which also shows the explicit
values calculated in Appendices \ref{app:harmonicoscillatormatrixelements} and
\ref{app:HOcoefficients} for an isotropic harmonic trap. We use the notation
that $\alpha_{m}^{(n)}$ and $\beta_{m}^{(n)}(\omega)$ are associated with
$n^{\text{th}}$-order, $m$-body processes. The sum that gives $\beta_{2}%
^{(2)}(\omega)$ diverges with cutoff as $\sqrt{\omega_{c}/\omega}$, where
$\omega_{c}/\omega$ is approximately the number of harmonic oscillator levels
included in the sum as a function of $\omega,$ for a fixed cutoff $\omega
_{c}.$ The coefficient $\alpha_{3}^{(2)}$ is convergent and in the limit
$\omega_{c}/\omega\rightarrow\infty$ is independent of $\omega.$ In the
following, we only indicate the explicit $\omega$ dependence for coefficients
that remain sensitive to $\omega$ in the limit $\omega_{c}/\omega
\rightarrow\infty,$ e.g., we write $\beta_{2}^{(2)}(\omega)$ but $\alpha
_{3}^{(2)}.$ We use a hard cutoff to numerically evaluate the coefficients
$\alpha_{3}^{(3)},\alpha_{4,1}^{(3)}$ and $\alpha_{4,1}^{(3)}$ (see
Sec.~\ref{sec:thirdorderenergies} for the definitions of the third-order
coefficients.) For the coefficients $\alpha_{3}^{(2)},\alpha_{4,3}^{(3)},$ and
$\alpha_{5}^{(3)},$ we find analytic results in the limit $\omega_{c}%
/\omega\rightarrow\infty.$ Finally, using the exponential regulator, we obtain
analytic results for the coefficients $\beta_{2}^{(2)}(\omega),\beta_{2}%
^{(3)}(\omega),$ and $\beta_{3}^{(3)}(\omega)$ for any $\omega_{c}/\omega.$

Equations (\ref{U22}) and (\ref{U32}) have also been represented
diagrammatically, with factors of $K_{ij;kl}[a_{\text{t}}(\omega_{0}%
)/\sigma(\omega)]$ assigned vertices
\raisebox{-0.05in}{\includegraphics[]{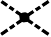}}, and contractions
$\contraction{}{\hat{a}}{_{a}}{\hat{a}}\hat{a}_{i}\hat{a}_{k}^{\dag}$
representing excited particles assigned dashed lines
$\raisebox{+0.02in}{\includegraphics[]{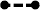}}.$ Uncontracted
operators $\hat{a}_{0}$ (or $\hat{a}_{0}^{\dag}$) are assigned incoming
\raisebox{+0.02in}{\includegraphics[]{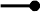}} (or outgoing
\raisebox{+0.02in}{\includegraphics[]{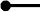}} ) lines. The counterterm is
represented as $K_{ij;kl}[a_{\text{ct}}(\omega_{0})/\sigma(\omega
)]=\raisebox{-0.05in}{\includegraphics[]{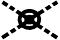}}.$ Intermediate
states have one or more excited particles and contribute an energy denominator
$1/\Delta\varepsilon_{ij}$. For example, the diagram
$\raisebox{-0.05in}{\includegraphics[]{beta1small}}$ is a graphical
representation for the term $\alpha_{3}^{(2)}[a_{\text{t}}(\omega_{0}%
)/\sigma(\omega)]^{2}$, and $U_{3}^{(2)}(\omega;\omega_{0}%
)=-6\raisebox{-0.05in}{\includegraphics[]{beta1small}}.$ We obtain
combinatorial prefactors [e.g.~$-6$ for $U_{3}^{(2)}\left(  \omega,\omega
_{0}\right)  $] from Wick's theorem by counting the number of equivalent
contractions, dividing by $2$ for every factor of $a_{\text{t}}(\omega_{0})$
or $a_{\text{ct}}(\omega_{0}),$ and multiplying by $m!$ for an $m$-body term.

The renormalization condition through second order is $U_{2}(\omega;\omega
_{0})=U_{2}^{(1)}(\omega_{0};\omega_{0})+U_{2}^{(2)}(\omega_{0};\omega
_{0})+\mathcal{O}([a_{\text{t}}(\omega_{0})]^{3})+\mathcal{O}(r_{\text{eff}%
}[a_{\text{t}}(\omega_{0})]^{2})=U_{2}^{(1)}(\omega_{0};\omega_{0}),$ and
hence $U_{2}^{(2)}(\omega_{0};\omega_{0})=0.$ Diagrammatically
\begin{equation}
\raisebox{-0.05in}{\includegraphics[]{counterterm1small}}|_{\omega=\omega
_{0}}=\raisebox{-0.05in}{\includegraphics[]{alpha1small}}|_{\omega=\omega_{0}%
}.
\end{equation}
Solving for the counterterm gives
\begin{equation}
a_{\text{ct}}(\omega_{0})=\frac{\beta_{2}^{(2)}(\omega_{0})}{\alpha_{2}^{(1)}%
}\left(  \frac{a_{\text{t}}(\omega_{0})}{\sigma(\omega_{0})}\right)
^{2}\sigma(\omega_{0}). \label{chi_omega0}%
\end{equation}
Substituting into Eq. (\ref{U22}) gives%
\begin{equation}
U_{2}^{(2)}(\omega;\omega_{0})=c_{2}^{(2)}(\omega,\omega_{0})\left(
\frac{a_{\text{t}}(\omega_{0})}{\sigma(\omega)}\right)  ^{2}, \label{U22b}%
\end{equation}
where the function
\begin{equation}
c_{2}^{(2)}(\omega,\omega_{0})=\sqrt{\omega_{0}/\omega}\beta_{2}^{(2)}%
(\omega_{0})-\beta_{2}^{(2)}(\omega) \label{c22func}%
\end{equation}
can be used for any $\omega.$ (We have used $\sigma(\omega_{0})/\sigma
(\omega)=\sqrt{\omega/\omega_{0}}$ above to simplify the expressions.)

The form of the expression for the coefficient $c_{2}^{(2)}\left(
\omega,\omega_{0}\right)  $ ensures that the divergent terms cancel. For an
isotropic harmonic oscillator, we show in App.~\ref{app:beta2(2)}, using an
exponential regulator, that%
\begin{equation}
\beta_{2}^{(2)}(\omega)=\left(  2/\pi\right)  [\sqrt{\omega_{c}/2\omega
}-(1-\log2)]+\mathcal{O}(1/\omega_{c}^{1/2}), \label{alpha2w}%
\end{equation}
and thus%
\begin{equation}
c_{2}^{(2)}(\omega,\omega_{0})=\left(  2/\pi\right)  (1-\log2)\left[
1-\sqrt{\omega_{0}/\omega}\right]  .
\end{equation}
The renormalization condition is automatically satisfied since $c_{2}%
^{(2)}(\omega_{0},\omega_{0})=0.$ For the special case when $\omega_{0}=0,$ we
find
\begin{equation}
c_{2}^{(2)}(\omega,0)=c_{2}^{(2)}=\left(  2/\pi\right)  (1-\log2)=0.19535....
\end{equation}
For brevity, we define $c_{m}^{(n)}$ without arguments as the coefficients
$c_{m}^{(n)}(\omega,0)$ for the special case when $\omega_{0}=0.$ In this
limit, the coefficients $c_{m}^{(n)}$ are independent of $\omega$.

Combining the first- and second-order contributions for the two-body
interaction energy gives
\begin{align}
U_{2}(\omega;\omega_{0}) & =c_{2}^{(1)}\left(  \frac{a_{\text{t}}(\omega_{0}%
)}{\sigma(\omega)}\right)  +c_{2}^{(2)}(\omega,\omega_{0})\left(
\frac{a_{\text{t}}(\omega_{0})}{\sigma(\omega)}\right)  ^{2} \nonumber \\ &+\mathcal{O}%
\left(  \frac{[a_{\text{t}}(\omega_{0})]^{3}}{[\sigma(\omega)]^{3}}\right)
+\mathcal{O}\left(  \frac{r_{\text{eff}}[a_{\text{t}}(\omega_{0})]^{2}%
}{[\sigma(\omega)]^{3}}\right)  .
\end{align}
The coefficient $\alpha_{3}^{(2)}$ in Eq.~(\ref{U32}) is finite and does not
require a regulator. For the three-body interaction energy we obtain%
\begin{align}
U_{3}(\omega;\omega_{0}) & =c_{3}^{(2)}\left(  \frac{a_{\text{t}}(\omega_{0}%
)}{\sigma(\omega)}\right)  ^{2}+\mathcal{O}\left(  \frac{[a_{\text{t}}%
(\omega_{0})]^{3}}{[\sigma(\omega)]^{3}}\right) \nonumber \\ & +\mathcal{O}\left(
\frac{r_{\text{eff}}[a_{\text{t}}(\omega_{0})]^{2}}{[\sigma(\omega)]^{3}%
}\right)  ,
\end{align}
where $c_{3}^{(2)}=-6\alpha_{3}^{(2)}=-0.85576...$ This value was previously
obtained in \cite{Johnson2009}, and is also calculated in
App.~\ref{app:alpha3(2)}.

\section{Effective interactions through third
order\label{sec:thirdorderenergies}}

\begin{figure*}[t]
\centering
\includegraphics[]{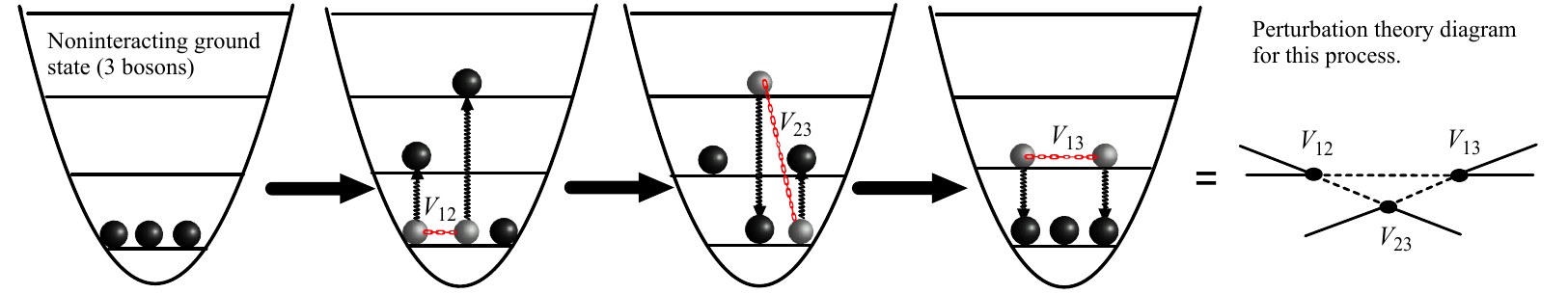} \caption{Sequence of boson-boson
interaction induced transitions to higher orbitals. This example generates
corrections to the ground state energy that can be viewed as an effective
three-body interaction. The process, which involves three interaction
vertices, arises at third order in perturbation theory and gives the energy
shift $\alpha_{3}^{(3)}[a_{\text{t}}(\omega_{0})/\sigma(\omega)]^{3}$ derived
in the text. Links labelled $V_{ij}$ represent intrinsic 2-body interactions
between particles $i$ and $j$. Black arrows represent virtual transitions to
and from excited orbitals. Solid and dashed lines represent atoms in ground
and excited vibrational states, respectively. The diagram on the far right
shows the perturbation theory diagram for this process.}%
\label{fig:examplesequence}%
\end{figure*}

We now extend our analysis to third order in the scattering length
$a_{\text{t}}(\omega_{0})$. This is necessary to obtain the leading-order
effective four-body interaction. Including the counterterm and effective-range
interaction, the formula for the third-order energy shift is
\begin{align}
& E^{(3)}(\omega;\omega_{0})=\sum_{ij\neq00,kl\neq00}^{\omega_{c}/\omega}%
\frac{V_{00,ij}V_{ij,kl}V_{kl,00}}{\Delta\varepsilon_{ij}\Delta\varepsilon
_{kl}} \label{E3} \\ &-V_{00,00}\sum_{ij\neq00}^{\omega_{c}/\omega}\frac{V_{00,ij}V_{ij,00}%
}{\Delta\varepsilon_{ij}^{2}}-2\sum_{ij\neq00}^{\omega_{c}/\omega}%
\frac{V_{\text{ct;}00,ij}V_{ij,00}}{\Delta\varepsilon_{ij}}+V_{00,00}^{\prime
}. \nonumber
\end{align}
The first term on the right-hand-side of Eq.~(\ref{E3}) gives
\begin{multline}
\frac{1}{8}\left(  \frac{a_{\text{t}}(\omega_{0})}{\sigma(\omega)}\right)
^{3}N(N-1) \times \\ \sum_{ij\neq00,klqr,st\neq00}^{\omega_{c}/\omega}\frac
{K_{00;ij}K_{kl;qr}K_{st;00}}{\Delta\varepsilon_{ij}\Delta\varepsilon_{st}%
}\langle\hat{a}_{i}\hat{a}_{j}\hat{a}_{k}^{\dagger}\hat{a}_{l}^{\dagger}%
\hat{a}_{q}\hat{a}_{r}\hat{a}_{s}^{\dagger}\hat{a}_{t}^{\dagger}\rangle,
\label{E3_1}%
\end{multline}
where the expectation value is with respect to the noninteracting ground state
with $(N-2)$ bosons. Applying Wick's theorem, Eq.~(\ref{E3_1}) expands as
\begin{flalign}
&\left(\frac{a_{\text{t}}(\omega_{0})}{\sigma(\omega)}\right)^{3} \times \Big(
\frac{1}{2!}\beta_{2}^{(3)}(\omega)N(N-1)& \label{ExpandingE3_1} \\& +\frac{1}{3!}%
[12\alpha_{3}^{(3)}+12\beta_{3}^{(3)}(\omega)]N(N-1)(N-2)& \nonumber\\
  &+\frac{1}{4!}[48\alpha_{4,1}^{(3)}+48\alpha_{4,2}^{(3)}+6\alpha_{4,3}%
^{(3)}]N(N-1)(N-2)(N-3)& \nonumber\\
  &+\frac{1}{5!}60\alpha_{5}^{(3)}N(N-1)(N-2)(N-3)(N-4)\Big).& \nonumber
\end{flalign}
We find effective two-, three-, and four-body interactions from the terms with
four, three, and two contractions, respectively. The zero-contraction term
vanishes since $\langle$:$\hat{a}_{i}\hat{a}_{j}\hat{a}_{k}^{\dagger}\hat
{a}_{l}^{\dagger}\hat{a}_{i^{\prime}}\hat{a}_{j^{\prime}}\hat{a}_{k^{\prime}%
}^{\dagger}\hat{a}_{l^{\prime}}^{\dagger}$:$\rangle=0.$ Comparing to
Eq.~(\ref{Eww0}), we see that there is a two-body contribution $\beta
_{2}^{(3)}(\omega)[a_{\text{t}}(\omega_{0})/\sigma(\omega)]^{3}%
=\raisebox{-0.0692in}{\includegraphics[]{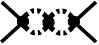}},$ there are two
three-body contributions $12\alpha_{3}^{(3)}[a_{\text{t}}(\omega_{0}%
)/\sigma(\omega)]^{3}=12\raisebox{-0.05in}{\includegraphics[]{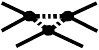}}$
and $12\beta_{3}^{(3)}(\omega)[a_{\text{t}}(\omega_{0})/\sigma(\omega
)]^{3}=12\raisebox{-0.05in}{\includegraphics[]{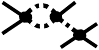}},$ and so on. The
definitions for the coefficients $\beta_{2}^{(3)},\alpha_{3}^{(3)},\beta
_{3}^{(3)},$ etc., are given in Table~\ref{table:alphaandbeta}, along with the
associated diagrams, asymptotic behavior, and explicit forms for an isotropic
harmonic oscillator potential (calculated in Appendices
\ref{app:harmonicoscillatormatrixelements} and \ref{app:HOcoefficients}).
Figure~\ref{fig:examplesequence} illustrates one of the sequences of virtual
transitions giving rise to $\alpha_{3}^{(3)}.$

\begin{table*}[tbh]
{\footnotesize \centering$%
\begin{tabular}
[c]{|ll||l|l|l|}\hline
Energies & (Diagrams) & Coefficients & Asymp. & Isotropic H.O. coefficients
($\omega_{c}\rightarrow\infty$)\\\hline\hline
\multicolumn{5}{|l|}{1$^{\text{st}}$-order in $\xi_{\text{t}}=a_{\text{t}%
}(\omega_{0})/\sigma(\omega)$}\\\hline\hline
$\alpha_{2}^{(1)}\xi_{\text{t}}$ &
=$\raisebox{-0.07in}{\includegraphics[]{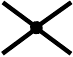}}$ & $\alpha_{2}^{(1)}%
=K_{0000}$ & N.A. & $\sqrt{\frac{2}{\pi}}=+0.797885...$\\\hline\hline
\multicolumn{5}{|l|}{2$^{\text{nd}}$-order in $\xi_{\text{t}}$}\\\hline\hline
$\alpha_{3}^{(2)}\xi_{\text{t}}^{2}$ &
=\ $\raisebox{-0.12in}{\includegraphics[]{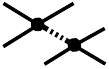}}$ & $\alpha_{3}^{(2)}%
=\sum\frac{K_{000i}K_{i000}}{\Delta\varepsilon_{i0}}$ & $a+e^{-\omega
_{c}/\omega}$ & $\left(  \frac{2}{\pi}\right)  [\frac{2\sqrt{3}}{3}%
+\log(8-4\sqrt{3})-1]=+0.142626...$\\\hline\hline
$\beta_{2}^{(2)}(\omega)\xi_{\text{t}}^{2}$ &
=\ $\raisebox{-0.07in}{\includegraphics[]{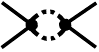}}$ & $\beta_{2}^{(2)}%
=\sum\frac{K_{00ij}K_{ij00}}{\Delta\varepsilon_{ij}}$ & $\sqrt{\frac
{\omega_{c}}{\omega}}$ & $\left(  \frac{2}{\pi}\right)  [\sqrt{\frac
{\omega_{c}}{2\omega}}-(1-\log2)-\frac{3}{2}\sqrt{\frac{\omega}{2\omega_{c}}%
}]$\\\hline\hline
\multicolumn{5}{|l|}{3$^{\text{rd}}$-order in $\xi_{\text{t}}$}\\\hline\hline
$\beta_{2}^{(3)}(\omega)\xi_{\text{t}}^{3}$ &
=\ $\raisebox{-0.07in}{\includegraphics[]{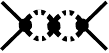}}$ & $\beta_{2}^{(3)}%
=\sum\frac{K_{00ij}K_{ijkl}K_{kl00}}{\Delta\varepsilon_{ij}\Delta
\varepsilon_{kl}}$ & $\left(  \frac{\omega_{c}}{\omega}\right)  $ &
$[\beta_{2}^{(2)}(\omega)]^{2}/\alpha_{2}^{(1)}$\\\hline
$\beta_{3}^{(3)}(\omega)\xi_{\text{t}}^{3}$ &
=\ $\raisebox{-0.12in}{\includegraphics[]{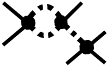}}$ & $\beta_{3}^{(3)}%
=\sum\frac{K_{00ij}K_{ij0k}K_{k000}}{\Delta\varepsilon_{ij}\Delta
\varepsilon_{k0}}$ & $\sqrt{\frac{\omega_{c}}{\omega}}$ & $\beta_{2}%
^{(2)}(\omega)\alpha_{3}^{(2)}/\alpha_{2}^{(1)}$\\\hline
$\alpha_{3}^{(3)}\xi_{\text{t}}^{3}$ &
=\ $\raisebox{-0.12in}{\includegraphics[]{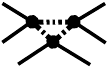}}$ & $\alpha_{3}^{(3)}%
=\sum\frac{K_{00ij}K_{j00k}K_{ik00}}{\Delta\varepsilon_{ij}\Delta
\varepsilon_{ik}}$ & $a+\sqrt{\frac{\omega}{\omega_{c}}}$ & $+0.56494\pm
0.00001\ $(estimate)\\\hline
$\alpha_{4,1}^{(3)}\xi_{\text{t}}^{3}$ &
=\ $\raisebox{-0.12in}{\includegraphics[]{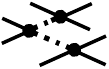}}$ & $\alpha_{4,1}^{(3)}%
=\sum\frac{K_{00ij}K_{j000}K_{i000}}{\Delta\varepsilon_{ij}\Delta
\varepsilon_{i0}}$ & $a+e^{-\omega_{c}/\omega}$ & $+0.077465...\ $%
(numerical)\\\hline
$\alpha_{4,2}^{(3)}\xi_{\text{t}}^{3}$ &
=\ $\raisebox{-0.12in}{\includegraphics[]{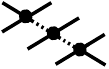}}$ & $\alpha_{4,2}^{(3)}%
=\sum\frac{K_{000i}K_{i00j}K_{j000}}{\Delta\varepsilon_{i0}\Delta
\varepsilon_{j0}}$ & $a+e^{-\omega_{c}/\omega}$ & $+0.051099...\ $%
(numerical)\\\hline
$\alpha_{4,3}^{(3)}\xi_{\text{t}}^{3}$ &
=\ $\raisebox{-0.12in}{\includegraphics[]{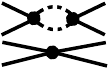}}$ & i$\alpha_{4,3}%
^{(3)}=\sum\frac{K_{00ij}K_{0000}K_{ij00}}{\Delta\varepsilon_{ij}^{2}}$ &
$a+\sqrt{\frac{\omega}{\omega_{c}}}$ & $\left(  \frac{2}{\pi}\right)
^{3/2}[\frac{\pi^{2}}{24}+\log2-\frac{1}{2}\left(  \log2\right)
^{2}]=+0.438946...$\\\hline
$\alpha_{5}^{(3)}\xi_{\text{t}}^{3}$ &
=\ $\raisebox{-0.12in}{\includegraphics[]{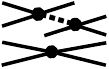}}$ & $\alpha_{5}^{(3)}%
=\sum\frac{K_{000i}K_{0000}K_{i000}}{\Delta\varepsilon_{i0}{}^{2}}$ &
$a+e^{-\omega_{c}/\omega}$ & $\frac{3}{4(2\pi)^{3/2}}{}_{4}F_{3}\left(
1,1,1,5/2;2,2,2;1/4\right)  =+0.051916...$\\\hline\hline
\multicolumn{5}{|c|}{Counterterms through third order}\\\hline
\multicolumn{5}{|l|}{$\chi_{\text{ct}}=\frac{a_{\text{ct}}(\omega_{0})}%
{\sigma(\omega)},$ $\ \alpha_{2}^{(1)}\chi_{\text{ct}}%
=\raisebox{-0.07in}{\includegraphics[]{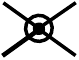}},\ \ \ \ \beta_{2}%
^{(2)}(\omega)\chi_{\text{ct}}\xi_{\text{t}}%
=\raisebox{-0.07in}{\includegraphics[]{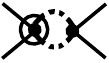}},\ \ \ \ \alpha_{3}%
^{(2)}\chi_{\text{ct}}\xi_{\text{t}}%
=\raisebox{-0.07in}{\includegraphics[]{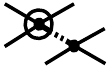}}$}\\\hline\hline
\multicolumn{5}{|c|}{Leading-order effective range terms}\\\hline
\multicolumn{5}{|l|}{$\alpha_{2}^{(1,2)}\left(  \frac{r_{\text{eff}}}%
{\sigma(\omega)}\right)  \xi_{\text{t}}^{2}%
=\raisebox{-0.05in}{\includegraphics[]{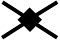}},$\ $\ \ \alpha
_{2}^{(1,2)}=K_{0000}^{\prime}=\frac{3}{4}\left(  \frac{2}{\pi}\right)
^{1/2}=+0.598413...$}\\\hline\hline
\multicolumn{5}{|c|}{Other relations:}\\\hline
\multicolumn{5}{|l|}{$\ \ \raisebox{-0.07in}{\includegraphics[]{gamma3}}$
(four-body) $=\raisebox{-0.07in}{\includegraphics[]{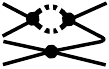}}$ (three-body)
$=\raisebox{-0.07in}{\includegraphics[]{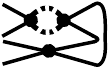}}$ (two-body) $=\alpha
_{4,3}^{(3)}\xi_{\text{t}}^{3}$}\\\hline
\multicolumn{5}{|l|}{$\ \ \raisebox{-0.07in}{\includegraphics[]{delta1}}$
(five-body)\ \ $=\raisebox{-0.07in}{\includegraphics[]{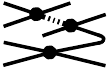}}$
(four-body)\ \ $=\raisebox{-0.07in}{\includegraphics[]{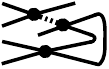}}$ (three-body)
$=\alpha_{5}^{(3)}\xi_{\text{t}}^{3}$}\\\hline
\end{tabular}
\ \ \ \ \ \ \ \ \ \ \ $ }\caption{The coefficients for all interaction
processes contributing to the two-, three-, and four-body interaction energies
through third-order in perturbation theory in $\xi_{\text{t}}=a_{\text{t}%
}(\omega_{0})/\sigma(\omega)$. The first column shows the diagrams from which
the $m$-body, $n^{\text{th}}$-order coefficients $\alpha_{m}^{(n)}$ and
$\beta_{m}^{(n)}(\omega)$ can be reconstructed. The coefficients as
multidimensional sums are given in the second column. Sums are over all
indices $i,j,k,...$ except combinations that give a zero energy term in the
denominator. The third column gives the asymptotic behavior of the
coefficients in terms of the cutoff $\omega_{c}$ and a constant $a$. The last
column gives the explicit values for the coefficients for an isotropic
harmonic oscillator potential. These values are obtained in the Appendices.
The table also shows the counterterm processes, the leading-order
effective-range contribution, and other relations needed for the renormalized
perturbation theory.}%
\label{table:alphaandbeta}%
\end{table*}

We next use Wick's theorem to evaluate the second term on the right-hand-side
of Eq.~(\ref{E3}), finding
\begin{align}
& \left[  -\frac{1}{4}\alpha_{4,3}^{(3)}N^{2}(N-1)^{2}-\frac{1}{2}\alpha
_{5}^{(3)}N^{2}(N-1)^{2}(N-2)\right] \nonumber \\ & \times \left(  \frac{a_{\text{t}}(\omega_{0}%
)}{\sigma(\omega)}\right)  ^{3}, \label{E32}%
\end{align}
where $\alpha_{4,3}^{(3)}$ and $\alpha_{5}^{(3)}$ already appear in
Eq.~(\ref{ExpandingE3_1}). Equation~(\ref{E32}) can be separated into $m$-body
contributions by expansion into terms proportional to $N(N-1),$ $N(N-1)(N-2),
$ etc.

It is surprising, at first sight, that $\alpha_{4,3}^{(3)}$ and $\alpha
_{5}^{(3)}$ contribute to several effective multi-body energies. From
Table~\ref{table:alphaandbeta}, $\alpha_{4,3}^{(3)}[a_{\text{t}}(\omega
_{0})/\sigma(\omega)]^{3}=\raisebox{-0.05in}{\includegraphics[]{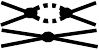}}$
and $\alpha_{5}^{(3)}[a_{\text{t}}(\omega_{0})/\sigma(\omega)]^{3}%
=\raisebox{-0.05in}{\includegraphics[]{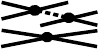}}$ look like processes
requiring four and five distinct particles, respectively. They also appear to
be composed of \textquotedblleft disconnected\textquotedblright\ sub-diagrams.
In RS perturbation theory, however, the second term in Eq.~(\ref{E3}) can be
reinterpreted in terms of particles going \textquotedblleft
backward\textquotedblright\ in time (right to left), or alternatively an
interpretation can be given in terms of holes. For example, the term
$-\alpha_{4,3}^{(3)}[a_{\text{t}}(\omega_{0})/\sigma(\omega)]^{3}$ gives a
two-body contribution if we view the two particles first going forward in time
(left to right), colliding to an excited intermediate state, colliding back to
the ground state, and finally going backward in time and colliding a third
time. Diagrammatically, this can be represented by the connected diagram
$\raisebox{-0.05in}{\includegraphics[]{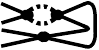}}.$ Similarly, if only one
particle goes back in time, it can collide with a third particle, giving the
three-body contribution $-6\alpha_{4,3}^{(3)}[a_{\text{t}}(\omega_{0}%
)/\sigma(\omega)]^{3}=-6\raisebox{-0.05in}{\includegraphics[]{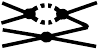}},$
which is also connected. In this paper, these related two-, three-, and
four-body diagrams have the same numerical value:
$\raisebox{-0.05in}{\includegraphics[]{alpha3small}}=\raisebox{-0.05in}{\includegraphics[]{beta3small}}=\raisebox{-0.05in}{\includegraphics[]{gamma3small}}.$%

The third term on the right-hand-side of Eq.~(\ref{E3}) gives the two- and
three-body counterterm contributions%
\begin{align}
&\left[  -\frac{1}{2}\beta_{2}^{(2)}(\omega)N(N-1)+\frac{1}{6}2\alpha_{3}%
^{(2)}N(N-1)(N-2)\right] \nonumber \\ & \times \left(  \frac{a_{\text{ct}}(\omega_{0})}%
{\sigma(\omega)}\right)  \left(  \frac{a_{\text{t}}(\omega_{0})}{\sigma
(\omega)}\right)  ,
\end{align}
or 
\begin{equation}
\beta_{2}^{(2)}(\omega)[a_{\text{t}}(\omega_{0})a_{\text{ct}}(\omega
_{0})/\sigma(\omega)^{2}%
]=\raisebox{-0.05in}{\includegraphics[]{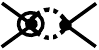}}
\end{equation}
and 
\begin{equation}
\alpha
_{3}^{(2)}[a_{\text{t}}(\omega_{0})a_{\text{ct}}(\omega_{0})/\sigma
(\omega)^{2}]=\raisebox{-0.05in}{\includegraphics[]{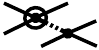}}.
\end{equation}
These counterterm contributions, shown in Table~\ref{table:alphaandbeta},
cancel the divergences from
$\raisebox{-0.05in}{\includegraphics[]{alpha1small}},$
$\raisebox{-0.05in}{\includegraphics[]{alpha2small}},$ and
$\raisebox{-0.05in}{\includegraphics[]{betactsmall}}.$ The disconnected
$5$-body contribution from Eq.~(\ref{E32}), generated by the term with a
single contraction, cancels with the disconnected five-body term in
Eq.~(\ref{ExpandingE3_1}), and there is no effective five-body interaction at
third order. Finally, the last term on the right-hand-side of Eq.~(\ref{E3})
gives the effective-range contribution%
\begin{equation}
E^{(1,2)}(\omega;\omega_{0})=\frac{1}{2}\alpha_{2}^{(1,2)}N(N-1)\left(
\frac{r_{\text{eff}}}{\sigma(\omega)}\right)  \left(  \frac{a_{\text{t}%
}(\omega_{0})}{\sigma(\omega)}\right)  ^{2},
\end{equation}
from which we extract the effective-range two-body interaction energy
\begin{equation}
U_{2}^{(1,2)}(\omega;\omega_{0})=\alpha_{2}^{(1,2)}\left(  \frac
{r_{\text{eff}}[a_{\text{t}}(\omega_{0})]^{2}}{[\sigma(\omega)]^{3}}\right)
=\raisebox{-0.05in}{\includegraphics[]{effectiverangesmall}}.
\end{equation}
The coefficient $\alpha_{2}^{(1,2)}\ $is given in
Table~\ref{table:alphaandbeta}. The special case $\omega_{0}=0$ gives
Eq.~($\ref{E12}$) and Eq.~(\ref{U2eff12}).

\subsection{Two-body interaction energy\label{sec:3rdorder2body}{}}

Adding all two-body contributions through third order, we obtain
\begin{align}
& U_{2}(\omega;\omega_{0}) =\raisebox{-0.05in}{\includegraphics[]{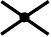}}-\raisebox{-0.05in}{\includegraphics[]{alpha1small}}+\raisebox{-0.05in}{\includegraphics[]{counterterm1small}}-\raisebox{-0.05in}{\includegraphics[]{alpha3small}}+\raisebox{-0.05in}{\includegraphics[]{alpha2small}}  \nonumber \\ & \hspace{15mm}-2\raisebox{-0.05in}{\includegraphics[]{counterterm2small}}+\raisebox{-0.05in}{\includegraphics[]{effectiverangesmall}}+\mathcal{O}%
(a_{\text{t}}^{4}) \\
&  =\alpha_{2}^{(1)}\left(  \frac{a_{\text{t}}(\omega_{0})}{\sigma(\omega
)}\right)  -\beta_{2}^{(2)}(\omega)\left(  \frac{a_{\text{t}}(\omega_{0}%
)}{\sigma(\omega)}\right)  ^{2} \nonumber \\ &+\alpha_{2}^{(1)}\left(  \frac{a_{\text{ct}%
}(\omega_{0})}{\sigma(\omega)}\right)  -\alpha_{4,3}^{(3)}\left(
\frac{a_{\text{t}}(\omega_{0})}{\sigma(\omega)}\right)  ^{3}\nonumber\\
&  +\beta_{2}^{(3)}(\omega)\left(  \frac{a_{\text{t}}(\omega_{0})}%
{\sigma(\omega)}\right)  ^{3}-2\beta_{2}^{(2)}(\omega)\left(  \frac
{a_{\text{ct}}(\omega_{0})}{\sigma(\omega)}\right)  \left(  \frac{a_{\text{t}%
}(\omega_{0})}{\sigma(\omega)}\right) \nonumber  \\ &+\alpha_{2}^{(1,2)}\left(
\frac{r_{\text{eff}}}{\sigma(\omega)}\right)  \left(  \frac{a_{\text{t}%
}(\omega_{0})}{\sigma(\omega)}\right)  ^{2} +\mathcal{O}(a_{\text{t}}^{4}). \label{U2thirdorder}%
\end{align}
Note that all diagrams in Eq.~(\ref{U2thirdorder}) are connected, when
interpreted in terms of both forward- and backward propagating particles. All
coefficients are given in Table~\ref{table:alphaandbeta}. For brevity, in
Eq.~(\ref{U2thirdorder}) and the following we adopt the convention that
$\mathcal{O}(a_{\text{t}}{}^{4})$ means $\mathcal{O}([a_{\text{t}}(\omega
_{0})/\sigma(\omega)]^{4})+\mathcal{O}(r_{\text{eff}}[a_{\text{t}}(\omega
_{0})]^{3}/\sigma(\omega)^{4}).$

The counterterm, found in the previous section to second order, must now be
recalculated using the renormalization condition through third order. This
adds a third-order term which cancels the divergence from $\beta_{2}%
^{(3)}(\omega)[a_{\text{t}}(\omega_{0})/\sigma(\omega)]^{3}%
=\raisebox{-0.0692in}{\includegraphics[]{alpha2small}},$ as well as the
effective range contribution $U_{2}^{(1,2)}(\omega;\omega_{0}).$ Solving the
renormalization condition $U_{2}(\omega;\omega_{0})=U_{2}^{(1)}(\omega
_{0};\omega_{0})+U_{2}^{(2)}(\omega_{0};\omega_{0})+U_{2}^{(3)}(\omega
_{0};\omega_{0})+U_{2}^{(1,2)}(\omega_{0};\omega_{0})+\mathcal{O}(a_{\text{t}%
}^{4})=U_{2}^{(1)}(\omega_{0};\omega_{0}),$ and hence $U_{2}^{(2)}(\omega
_{0};\omega_{0})+U_{2}^{(3)}(\omega_{0};\omega_{0})+U_{2}^{(1,2)}(\omega
_{0};\omega_{0})=0,$ we find $a_{\text{ct}}(\omega_{0})$ from
\begin{align}
&\alpha_{2}^{(1)}\left(  \frac{a_{\text{ct}}(\omega_{0})}{\sigma(\omega_{0}%
)}\right)  =\beta_{2}^{(2)}(\omega_{0})\left(  \frac{a_{\text{t}}%
(\omega_{0})}{\sigma(\omega_{0})}\right)  ^{2} \nonumber \\ & -\left[  \beta_{2}^{(3)}%
(\omega_{0})-2[\beta_{2}^{(2)}(\omega_{0})]^{2}/\alpha_{2}^{(1)}-\alpha
_{4,3}^{(3)}\right]  \left(  \frac{a_{\text{t}}(\omega_{0})}{\sigma(\omega
_{0})}\right)  ^{3}\nonumber\\
&  -d_{2}^{(1,2)}\left(  \frac{r_{\text{eff}}}{\sigma(\omega_{0})}\right)
\left(  \frac{a_{\text{t}}(\omega_{0})}{\sigma(\omega_{0})}\right)
^{2}+\mathcal{O}(a_{\text{t}}^{4}). \label{3rdordercounterterm}%
\end{align}
Diagrammatically, this can be expressed as
\begin{equation}\raisebox{-0.05in}{\includegraphics[]{counterterm1small}}-2\raisebox{-0.05in}{\includegraphics[]{counterterm2small}}=\raisebox{-0.05in}{\includegraphics[]{alpha1small}}+\raisebox{-0.05in}{\includegraphics[]{alpha3small}}-\raisebox{-0.05in}{\includegraphics[]{alpha2small}}-\raisebox{-0.05in}{\includegraphics[]{effectiverangesmall}},
\end{equation}
with all diagrams evaluated at $\omega=\omega_{0}.$ By including
$U_{2}^{(1,2)}(\omega;\omega_{0})$ in the counterterm equation, the
renormalization condition for $a_{\text{t}}(\omega_{0})$ includes both
zero-range and effective-range contributions. If we do not include
$U_{2}^{(1,2)}(\omega;\omega_{0})$ in the renormalization condition, then
$a_{\text{t}}(\omega_{0})$ is the trap scattering length for zero-range
potentials. Substituting the counterterm from Eq.~(\ref{3rdordercounterterm})
into Eq.~(\ref{U2thirdorder}) and using $\alpha_{2}^{(1)}\beta_{2}%
^{(3)}(\omega)=[\beta_{2}^{(2)}(\omega)]^{2},$ which is proven in
Appendix~\ref{app:beta2(3)}, we find after some algebra that
\begin{align}
&U_{2}(\omega;\omega_{0})    =c_{2}^{(1)}\left(  \frac{a_{\text{t}}(\omega
_{0})}{\sigma(\omega)}\right)  +c_{2}^{(2)}(\omega,\omega_{0})\left(
\frac{a_{\text{t}}(\omega_{0})}{\sigma(\omega)}\right)  ^{2} \nonumber \\ &+c_{2}%
^{(3)}(\omega,\omega_{0})\left(  \frac{a_{\text{t}}(\omega_{0})}{\sigma
(\omega)}\right)  ^{3} \nonumber \\ & +d_{2}^{(1,2)}(\omega,\omega_{0})\left(  \frac{r_{\text{eff}}}%
{\sigma(\omega)}\right)  \left(  \frac{a_{\text{t}}(\omega_{0})}{\sigma
(\omega)}\right)  ^{2}+\mathcal{O}(a_{\text{t}}^{4}), \label{U2}%
\end{align}
where $c_{2}^{(3)}(\omega,\omega_{0})$ and $d_{2}^{(1,2)}(\omega,\omega_{0})$
are given in Table~\ref{Table:cnmfunctions}.

Recall that in the formula $U_{2}(\omega;\omega_{0})$ the first argument
$\omega$ is the trap frequency for which we are interested in predicting the
two-body energy, and the second argument $\omega_{0}$ is the trap frequency at
which the two-body trap scattering length $a_{\text{t}}(\omega_{0})$ is
defined or measured. The coefficients for $\omega_{0}=0$ are given in
Table~\ref{Table:CoefficientsCnm}. If $r_{\text{eff}}=0,$ these values
reproduce through third order the exact solution for the ground state of two
harmonically trapped bosons with zero-range interactions found in
\cite{Busch1998}. This agreement between the quantum mechanical and quantum
field theory solutions is a nice illustration of how the renormalized
effective field theory captures the correct low-energy physics. Interestingly,
if $r_{\text{eff}}\neq0,$ our result for $U_{2}(\omega,0)$ still agrees with
the solution in \cite{Busch1998}, if that solution is Taylor expanded in
$a_{\text{t}}(0)$ and $r_{\text{eff}}[a_{\text{t}}(0)]^{2}$ after making the
substitution $a_{\text{f}}(0)\rightarrow a_{\text{f}}(0)+(1/2)r_{\text{eff}%
}[a_{\text{f}}(0)]^{2}k_{\text{rel}}^{2},$ providing further evidence of the
universality of the higher-order perturbative results derived here.

Another important special case is $\omega=\omega_{0}.$ Since $c_{2}%
^{(2)}(\omega_{0},\omega_{0})=c_{2}^{(3)}(\omega_{0},\omega_{0})=d_{2}%
^{(1,2)}(\omega_{0},\omega_{0})=0,$ the predicted two-body energy is%
\begin{equation}
U_{2}(\omega=\omega_{0};\omega_{0})=c_{2}^{(1)}\left(  \frac{a_{\text{t}%
}(\omega_{0})}{\sigma(\omega_{0})}\right)  +\mathcal{O}(a_{\text{t}}^{4}),
\end{equation}
reproducing the renormalization condition that $a_{\text{t}}(\omega_{0})$ is
the \emph{physical} trap scattering length for two bosons at frequency
$\omega_{0}.$

\begin{table}[h]
{\footnotesize \centering$%
\begin{tabular}
[c]{l}%
Frequency-Dependent Effective Interaction Coefficients\\\hline\hline
\ \ \ \ \emph{Two-body}\\
$c_{2}^{(1)}(\omega,\omega_{0})=\alpha_{2}^{(1)}=\left(  2/\pi\right)  ^{1/2}%
$\\
$c_{2}^{(2)}(\omega,\omega_{0})=\left(  2/\pi\right)  (1-\log2)\left[
1-\sqrt{\omega_{0}/\omega}\right]  $\\
$c_{2}^{(3)}(\omega,\omega_{0})=\left(  2/\pi\right)  ^{3/2}(1-\log
2)^{2}\left[  1-\sqrt{\omega_{0}/\omega}\right]  ^{2}$\\
$\ \ \ \ \ \ \ \ \ \ \ \ \ \ \ -\left(  2/\pi\right)  ^{3/2}\left(  \pi
^{2}/24+\log2-\frac{1}{2}\log^{2}2\right)  \left[  1-\omega_{0}/\omega\right]
$\\
$d_{2}^{(1,2)}(\omega,\omega_{0})=\alpha_{2}^{(1,2)}\left[  1-\omega
_{0}/\omega\right]  =(3/4)\left(  2/\pi\right)  ^{1/2}\left[  1-\omega
_{0}/\omega\right]  $\\
\ \ \ \ \emph{Three-body}\\
$c_{3}^{(2)}(\omega,\omega_{0})=-6\alpha_{3}^{(2)}$\\
$c_{3}^{(3)}(\omega,\omega_{0})=-12\alpha_{3}^{(2)}c_{2}^{(2)}(\omega
,\omega_{0})/\alpha_{2}^{(1)}$\\
$\ \ \ \ \ \ \ \ \ \ \ \ \ \ \ +[12\alpha_{3}^{(3)}-6\alpha_{4,3}%
^{(3)}-18\alpha_{5}^{(3)}]$\\
\ \ \ \ \emph{Four-body}\\
$c_{4}^{(3)}(\omega,\omega_{0})=48\alpha_{4,1}^{(3)}+48\alpha_{4,2}%
^{(3)}-72\alpha_{5}^{(3)}$%
\end{tabular}
\ \ \ \ \ \ \ \ \ \ \ \ \ $ }\caption{The functions $c_{m}^{(n)}(\omega
,\omega_{0}),$ which determine the $n^{\text{th}}$-order contributions to the
$m$-body effective interaction energies$,$ and $d_{2}^{(1,2)}(\omega
,\omega_{0}),$ which determines the leading-order effective-range correction,
for neutral bosons in a harmonic potential of frequency $\omega$, in terms of
the scattering length $a_{\text{t}}(\omega_{0})$ defined at trap frequency
$\omega_{0}.$ The special case $\omega_{0}=0$ reduces to the results given in
Table 1.}%
\label{Table:cnmfunctions}%
\end{table}

\subsection{\textbf{Three-body interaction energy\label{sec:3rdorder3body}}}

In \cite{Johnson2009}, we obtained the effective three-body interaction energy
to second order. We now determine the next-order correction by combining all
three-body contributions through third order, giving
\begin{flalign}
&U_{3}(\omega;\omega_{0}) =-6\raisebox{-0.05in}{\includegraphics[]{beta1small}}+12\raisebox{-0.05in}{\includegraphics[]{beta2small}}+12\raisebox{-0.05in}{\includegraphics[]{betactsmall}}-12\raisebox{-0.05in}{\includegraphics[]{counterterm3small}}& \nonumber \\ & \hspace{15mm}-6\raisebox{-0.05in}{\includegraphics[]{beta3small}}-18\raisebox{-0.05in}{\includegraphics[]{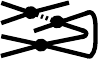}}+\mathcal{O}%
(a_{\text{t}}^{4})\label{U3}\\
&  =-6\alpha_{3}^{(2)}\left(  \frac{a_{\text{t}}(\omega_{0})}{\sigma(\omega
)}\right)  ^{2}+12\alpha_{3}^{(3)}\left(  \frac{a_{\text{t}}(\omega_{0}%
)}{\sigma(\omega)}\right)  ^{3} \nonumber \\ &+12\beta_{3}^{(3)}(\omega)\left(
\frac{a_{\text{t}}(\omega_{0})}{\sigma(\omega)}\right)  ^{3} -12\alpha_{3}^{(2)}\left(  \frac{a_{\text{ct}}(\omega_{0})}{\sigma(\omega
)}\right)  \left(  \frac{a_{\text{t}}(\omega_{0})}{\sigma(\omega)}\right) \nonumber \\ &
-6\alpha_{4,3}^{(3)}\left(  \frac{a_{\text{t}}(\omega_{0})}{\sigma(\omega
)}\right)  ^{3}-18\alpha_{5}^{(3)}\left(  \frac{a_{\text{t}}(\omega_{0}%
)}{\sigma(\omega)}\right)  ^{3}+\mathcal{O}(a_{\text{t}}^{4}).\nonumber
\end{flalign}
Representing the three-body contributions from $\alpha_{4,3}^{(3)}$ and
$\alpha_{5}^{(3)}$ using reversed (left to right) particle lines, as
previously described, we again find that only connected diagrams contribute.

For the three-body energy, it is sufficient to use the second-order
counterterm in Eq. (\ref{chi_omega0}). In Appendix~\ref{app:beta3(3)}, we show
that $\beta_{3}^{(3)}(\omega)=\beta_{2}^{(2)}(\omega)\alpha_{3}^{(2)}%
/\alpha_{2}^{(1)}$. From these results it follows that the difference between
the individually divergent contributions in Eq.~(\ref{U3}),
$12\raisebox{-0.05in}{\includegraphics[]{betactsmall}}$ and
$12\raisebox{-0.05in}{\includegraphics[]{counterterm3small}},$ is finite.
After some algebra, we find that%
\begin{align}
U_{3}(\omega;\omega_{0})= &c_{3}^{(2)}\left(  \frac{a_{\text{t}}(\omega_{0}%
)}{\sigma(\omega)}\right)  ^{2}+c_{3}^{(3)}(\omega,\omega_{0})\left(
\frac{a_{\text{t}}(\omega_{0})}{\sigma(\omega)}\right)  ^{3} \nonumber \\ &+\mathcal{O}%
(a_{\text{t}}^{4}),
\end{align}
where $c_{3}^{(2)}$ and $c_{3}^{(3)}(\omega,\omega_{0})$ are given in
Table~\ref{Table:cnmfunctions}.

If $\omega_{0}$ equals zero, we find
\begin{align}
c_{3}^{(3)} & =-12(1-\log2)\alpha_{2}^{(1)}\alpha_{3}^{(2)}+[12\alpha_{3}%
^{(3)}-6\alpha_{4,3}^{(3)}-18\alpha_{5}^{(3)}] \nonumber \\ &=+2.7921\pm0.0001.
\end{align}
The error reflects a one standard deviation uncertainty due to the
extrapolation of the numerical estimate for $\alpha_{3}^{(3)}$ to the limit
$\omega_{c}\rightarrow\infty\ $(see App.~\ref{app:alpha3(33)}). Another
special case is $\omega=\omega_{0},$ giving%
\begin{align}
c_{3}^{(3)}(\omega,\omega) =12\alpha_{3}^{(3)}-6\alpha_{4,3}^{(3)}-18\alpha
_{5}^{(3)} =+3.2112\pm0.0001.
\end{align}

\subsection{Four-body interaction energy\label{sec:3rdorder4body}}

Finally, we calculate the leading order contribution to the effective
four-body interaction energy. We find
\begin{align}
U_{4}(\omega;\omega_{0})  &
=48\raisebox{-0.05in}{\includegraphics[]{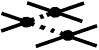}}+48\raisebox{-0.05in}{\includegraphics[]{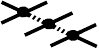}}-72\raisebox{-0.05in}{\includegraphics[]{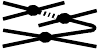}}+\mathcal{O}%
(a_{\text{t}}^{4})\nonumber\\
&  =c_{4}^{(3)}\left(  \frac{a_{\text{t}}(\omega_{0})}{\sigma(\omega)}\right)
^{3}+\mathcal{O}(a_{\text{t}}^{4}), \label{U4}%
\end{align}
with coefficient
\begin{equation}
c_{4}^{(3)}=48\alpha_{4,1}^{(3)}+48\alpha_{4,2}^{(3)}-72\alpha_{5}%
^{(3)}=+2.43317....
\end{equation}
As anticipated, the two disconnected terms that depend on $\alpha_{4,3}%
^{(3)}=\raisebox{-0.069in}{\includegraphics[]{gamma3small}}$ cancel and, at
this order, $c_{4}^{(3)}$ is independent of $\omega$ and $\omega_{0}.$ The
leading-order contribution to the four-body energy does not require
renormalization, as is true for all leading-order $m$-body terms. Comparison
of $c_{4}^{(3)}$ and $c_{3}^{(3)}$ reveals, however, that they are of similar
magnitude and therefore for a consistent and accurate treatment both
corrections need to be included. Because $c_{3}^{(3)}$ requires
renormalization, we see why the systematic renormalization of divergences is
needed even though the leading-order contribution to the four-body interaction
energy could be obtained without these considerations.

\section{Summary\label{sec:conclusion}}

We have derived effective two-, three-, and four-body interaction energies for
$N$ bosons in an isotropic harmonic trap of frequency $\omega$. These energies
are functions of the trap scattering length $a_{\text{t}}(\omega_{0})$ and
harmonic oscillator length $\sigma(\omega),$ and include both renormalization
effects due to quantum fluctuations to higher-orbitals and leading-order
finite-range corrections. The frequency $\omega_{0}$ at which the scattering
length is defined plays a role closely analogous to the low-energy scale at
which coupling constants are defined in high-energy effective field theories
(e.g., see~\cite{Srednicki2007}). The formulas for the interaction energies
are given in Eqs.~(\ref{U2}), (\ref{U3}), and(\ref{U4}), and are expressed in
terms of the functions $c_{m}^{(n)}(\omega;\omega_{0})$ given in
Table~\ref{Table:cnmfunctions}. In turn, these functions require the
coefficients $\alpha_{m}^{(n)}$ and $\beta_{m}^{(n)}(\omega)$ given in
Table~\ref{table:alphaandbeta}. The special case when $\omega_{0}=0$ is
summarized in Table~\ref{Table:CoefficientsCnm} and Eqs.~(\ref{U2w}),
(\ref{U3w}), and (\ref{U4w}). In Sec.~\ref{sec:comparisonofenergies}, we
showed that these results give excellent agreement to numerical simulations
for ultracold bosons interacting through a Gaussian model potential.

We find at third-order in $a_{\text{t}}(\omega_{0})$ that the shifts to the
effective three- and four-body interaction energies are comparable, showing
that the renormalized three-body interaction needs to be taken into account
when the leading-order four-body interactions are considered. In the future,
we plan to use this formalism to determine the effective multi-body
interactions for other potentials, such as anisotropic traps or the anharmonic
sites of an optical lattice. The cross-over from the perturbative small
scattering length regime to the universal regime of Efimov physics is also
very interesting and diagrammatic resummation techniques can be used to study
the onset of nonperturbative behaviors. For example, collapse and revival
experiments suggest that four- and higher-body interactions may be present in
the data \cite{Will2010}, but our results also show that in these systems
$a_{\text{t}}(0)/\sigma(\omega_{0})$ is large enough for significant
nonperturbative effects to be important, and we would like to better
understand this physics within our framework. A unified description of elastic
and inelastic interactions (e.g.~three-body recombination physics
\cite{Burt1997,Esry1999}) would also be useful.

More immediately, the results in this paper can be applied to investigations
of finite-range interactions, can be used for precision experiments probing
for the possible existence of intrinsic three- and higher-body interactions,
and can enable explorations of fundamental concepts in effective field theory
including renormalization and energy-dependent (running) coupling constants.
For example, the influence of intrinsic higher-body interactions would cause
deviations from our predictions, which are based on only intrinsic two-body
interactions. Moreover, we can engineer and exploit useful effective
interactions using a combination of magnetic Feshbach resonances
\cite{Chin2010} and the ability to tune the few-body interactions by
controlling the trap parameters and shape \cite{Bolda2002,Blume2007}. One of
our longer-term goals is to use this physics to develop nonlinear measurement
techniques. For example, the nonlinear dynamics seen in collapse-and-revival
experiments can lead to better than shot-noise measurements of the $m$-body
interaction energies, or it may be possible to exploit strongly correlated
non-equilibrium states in lattices for new types of sensing. In this way, the
rich physics of renormalization and nonlinear quantum dynamics could be used
to create new types ultra-cold atom simulators, quantum information
processors, or quantum sensors.

\section{Acknowledgements\label{sec:acknowledgements}}

PRJ and ET acknowledge support from the U.S. Army Research Office under
contract/grant 60661PH. PRJ acknowledges additional support from the Research
Corporation for Science Advancement and computing resources provided by the
American University High Performance Computing System. ET acknowledges support
from a NSF Physical Frontier Center. DB and XYY gratefully acknowledge
fruitful discussions with Kevin Daily and support by the NSF through grant
PHY-0855332. PRJ thanks Nathan Harshman for helpful discussions.

\appendix

\section{$\delta$-function boson-boson interaction matrix elements for an
isotropic harmonic trap\label{app:harmonicoscillatormatrixelements}}

This appendix derives interaction matrix elements for bosons in an isotropic
harmonic oscillator trap with frequency $\omega$ and zero-range $\delta
$-function interactions. Alternative methods for obtaining these matrix
elements are given in \cite{talman_properties_1970,Edwards1996}.

\subsection{Isotropic harmonic oscillator
wavefunctions\label{app:wavefunctions}}

The calculations are most conveniently performed in coordinates scaled by the
harmonic oscillator length $\sigma(\omega).$ In spherical coordinates, the
normalized, dimensionless isotropic harmonic oscillator states $|nlm\rangle$
have wavefunctions $\phi_{nlm}\left(  \mathbf{r}\right)  =\langle
\mathbf{r}|nlm\rangle=\chi_{nl}\left(  r\right)  Y_{lm}\left(  \theta
,\phi\right)  $, where $Y_{lm}\left(  \theta,\phi\right)  $ are spherical
harmonics$.$ The radial functions are
\begin{equation}
\chi_{nl}\left(  r\right)  =N_{nl}r^{l}e^{-r^{2}/2}L_{n}^{(l+1/2)}\left(
r^{2}\right)  ,
\end{equation}
where $L_{n}^{(\alpha)}\left(  r\right)  $ are associated Laguerre
polynomials,%
\begin{equation}
N_{nl}=\sqrt{\frac{2\Gamma(n+1)}{\Gamma(n+l+3/2)}}%
\end{equation}
are normalization constants, and
\begin{equation}
L_{n}^{(l+1/2)}\left(  0\right)  =\frac{\Gamma\left(  n+l+3/2\right)  }%
{\Gamma\left(  n+1\right)  \Gamma\left(  l+3/2\right)  }.
\end{equation}
The single-particle ground state is $\phi_{000}(\mathbf{r})=\pi^{-3/4}%
e^{-r^{2}/2}.$ Recall that we use the shorthand notation $i=\{nlm\}$ for
states with vibrational quantum number $n,$ angular momentum $l,$ and angular
momentum projection quantum number $m.$ The single-particle energies are
$\varepsilon_{i}=\varepsilon_{nlm}=2n+l+3/2.$ A complete set of
(un-symmetrized) two-particle wavefunctions is $|ij\rangle=|n_{1}l_{1}%
m_{1},n_{2}l_{2}m_{2}\rangle.$ For convenience, we define the (dimensionless)
two-particle energy differences%
\begin{align}
\Delta\varepsilon_{ij}  &  \equiv\Delta\varepsilon_{n_{1}l_{1}m_{1},n_{2}%
l_{2}m_{2}}=\varepsilon_{n_{1}l_{1}m_{1}}+\varepsilon_{n_{2}l_{2}m_{2}%
}-2\varepsilon_{000}\\
&  =2n_{1}+2n_{2}+l_{1}+l_{2}. \label{EnergyDiffs}%
\end{align}

\subsection{Matrix elements in the single-particle
basis\label{app:spmatrixelements}}

The matrix elements $K_{ij;kl}$ defined in Eq.~(\ref{matrix elements})
correspond to transitions $|kl\rangle\rightarrow|ij\rangle$ with two-boson
basis functions $|ij\rangle$ and $|kl\rangle$ from the $|n_{1}l_{1}m_{1}%
,n_{2}l_{2}m_{2}\rangle$ basis. In this subsection, we evaluate the subset
$K_{ij;00}$ of these matrix elements given by
\begin{align}
& K_{n_{1}l_{1}m_{1},n_{2}l_{2}m_{2};000,000}  \nonumber \\ &  =4\pi\int\phi_{n_{1}l_{1}%
m_{1}}^{\ast}\left(  \mathbf{r}\right)  \phi_{n_{2}l_{2}m_{2}}^{\ast}\left(
\mathbf{r}\right)  \phi_{000}\left(  \mathbf{r}\right)  \phi_{000}\left(
\mathbf{r}\right)  d\mathbf{r} \nonumber \\
&  =\delta_{l_{1},l_{2}}\delta_{m_{1},-m_{2}}K_{\text{s.p.}}(n_{1},n_{2}%
,l_{1}), \label{deltaKsp}%
\end{align}
where $\delta_{a,b}$ is the Kronecker-delta and
\begin{align}
&K_{\text{s.p.}}(n_{1},n_{2},l) =\frac{4}{\sqrt{\pi}}N_{n_{1}l}N_{n_{2}l} \nonumber \\ & \times
\int L_{n_{1}}^{(l+1/2)}\left(  r^{2}\right)  L_{n_{2}}^{(l+1/2)}\left(
r^{2}\right)  e^{-2r^{2}}r^{2l+2}dr. \label{Kspintegral}%
\end{align}
The subscript \textquotedblleft s.p.\textquotedblright\ means single-particle
basis, and we have used the orthonormality of the spherical harmonics.

We next use the complex contour integral representation \cite{DLMF2011}
\begin{equation}
L_{n}^{(l+1/2)}\left(  r^{2}\right)  =\frac{1}{2\pi i}\oint\frac
{e^{-r^{2}z/\left(  1-z\right)  }}{\left(  1-z\right)  ^{l+3/2}z^{n+1}}dz,
\end{equation}
with a clockwise contour circling the pole at $z=0.$ Substituting and then
integrating over $r$ gives
\begin{align}
& K_{\text{s.p.}}(n_{1},n_{2},l)=\frac{2}{\sqrt{\pi}}N_{n_{1}l}N_{n_{2}l}%
\Gamma(l+3/2) \times \nonumber \\ & \frac{1}{2\pi i}\oint\frac{dz_{1}}{z_{1}^{n_{1}+1}}\frac{1}{2\pi
i}\oint\frac{dz_{2}}{z_{2}^{n_{2}+1}}\frac{1}{\left(  2-z_{1}-z_{2}\right)
^{l+3/2}}.
\end{align}
Applying the Cauchy residue theorem twice, first integrating counter-clockwise
around the pole at $z_{2}=0,$ and then around $z_{1}=0,$ and substituting in
the expressions for the normalization constants $N_{nl}$ gives
\begin{align}
& K_{\text{s.p.}}(n_{1},n_{2},l) =\sqrt{\frac{2}{\pi}} \times \label{Kn1n2l} \\ & \frac{2^{-n_{1}-n_{2}%
-l}\Gamma(n_{1}+n_{2}+l+3/2)}{\sqrt{\Gamma(n_{1}+1)\Gamma(n_{2}+1)\Gamma
(n_{1}+l+3/2)\Gamma(n_{2}+l+3/2)}}. \nonumber %
\end{align}
This expression also gives the matrix element for the transition
$|0i\rangle\rightarrow|0k\rangle.$

\subsection{Matrix elements in relative and center-of-mass particle basis
\label{app:relmatrixelements}}

It is simpler to compute some matrix elements by switching to a basis of
states $|\tilde{\imath}\tilde{j}\rangle=|nlm,NLM\rangle,$ with normalized
relative and center-of-mass wavefunctions $\tilde{\phi}_{nlm}(\mathbf{r)}%
\tilde{\Phi}_{NLM}(\mathbf{R)}$ defined in terms of coordinates $\mathbf{r=(r}%
_{1}-\mathbf{r}_{2})/\sqrt{2}$ and $\mathbf{R=(r}_{1}+\mathbf{r}_{2})/\sqrt
{2},$ and (dimensionless) two-particle energy differences%
\begin{equation}
\Delta\varepsilon_{nlm,NLM}\equiv2n+l+2N+L.
\end{equation}

Working in the $|nlm,NLM\rangle$ basis and using the fact that the
interactions conserve the center-of-mass motion, the matrix elements for the
transitions $|\tilde{k}\tilde{l}\rangle\rightarrow|\tilde{\imath}\tilde
{j}\rangle$ are $K_{\tilde{\imath}\tilde{j};\tilde{k}\tilde{l}}%
=K_{nlm,n^{\prime}l^{\prime}m^{\prime};NLM,N^{\prime}L^{\prime}M^{\prime}%
}=K_{\text{rel}}(n,n^{\prime})\delta_{l,0}\delta_{m,0}\delta_{l^{\prime}%
,0}\delta_{m^{\prime},0}\delta_{N,N^{\prime}}\delta_{L,L^{\prime}}%
\delta_{M,M^{\prime}},$ where
\begin{equation}
K_{\text{rel}}(n,n^{\prime})=\sqrt{\frac{2}{\pi}}\frac{\tilde{\phi}%
_{n00}^{\ast}(\mathbf{0)}\tilde{\phi}_{n^{\prime}00}(\mathbf{0)}}%
{\mathbf{|}\tilde{\phi}_{000}(\mathbf{0)|}^{2}} \label{Knlmnlm}%
\end{equation}
only depends on the principle quantum numbers for the relative motion. Below
we use the fact that $K_{\text{rel}}(n,n^{\prime})$ factors as%
\begin{equation}
K_{\text{rel}}(n,n^{\prime})=\sqrt{\frac{\pi}{2}}K_{\text{rel}}%
(n,0)K_{\text{rel}}(n^{\prime},0), \label{Krelnn}%
\end{equation}
and%
\begin{equation}
K_{\text{rel}}(n,0)=\frac{2}{\pi^{3/4}}\sqrt{\frac{\Gamma(n+3/2)}{\Gamma
(n+1)}}. \label{Kreln}%
\end{equation}
Also, $K_{\text{rel}}(0,0)\ $equals$\sqrt{2/\pi}.$

\section{Perturbation theory coefficients through third
order\label{app:HOcoefficients}}

In this appendix, we compute for neutral bosons in an isotropic harmonic
potential the $m$-body, $n^{\text{th}}$-order coefficients $\alpha_{m}^{(n)}$
and $\beta_{m}^{(n)}(\omega)$ needed for the perturbation theory through third
order. We first evaluate the coefficients $\alpha_{3}^{(2)},\alpha_{3}%
^{(3)},\alpha_{4,1}^{(3)},\alpha_{4,2}^{(3)},\alpha_{4,3}^{(3)},$ and
$\alpha_{5}^{(3)},$ which are finite and $\omega$-independent in the limit
that $\omega_{c}/\omega\rightarrow\infty.$ Then we evaluate the coefficients
$\beta_{2}^{(2)}(\omega),\beta_{2}^{(3)}(\omega),$ and $\beta_{3}^{(3)}%
(\omega),$ which diverge as $\omega_{c}/\omega\rightarrow\infty.$

\subsection{Three-body, second-order coefficient $\alpha_{3}^{(2)}%
$\label{app:alpha3(2)}}

In the single-particle basis $|n_{1}l_{1}m_{1},n_{2}l_{2}m_{2}\rangle,$ the
contribution $\raisebox{-0.05in}{\includegraphics[]{beta1small}}$ has the
coefficient%
\begin{equation}
\alpha_{3}^{(2)}=\sum\limits_{i\neq0}\frac{K_{00;0i}K_{i0;00}}{\Delta
\varepsilon_{i0}},
\end{equation}
where the sum $\sum_{i\neq0}$ is over all allowed single-particle states
excluding the ground state. Due to angular momentum conservation only
$i=\{nlm\}$ with $l=m=0$ contribute, and $\Delta\varepsilon_{i0}=2n$.
Evaluating the sum gives the analytic result
\begin{align}
&\alpha_{3}^{(2)}=\sum_{n=1}^{\infty}\frac{K_{\text{s.p.}}(n,0,0)^{2}}%
{2n}  \\ &=\left(  \frac{2}{\pi}\right)  \left(  \frac{2\sqrt{3}}{3}+\log
(8-4\sqrt{3})-1\right) =0.142626....\nonumber
\end{align}
The terms in the summand become smaller exponentially with $n$, and
$\alpha_{3}^{(2)}$ converges to the asymptotic form $a+\mathcal{O}%
(e^{-\omega_{c}/\omega}).$ This behavior will be true for all
\textquotedblleft tree diagrams\textquotedblright\ which, like
$\raisebox{-0.05in}{\includegraphics[]{beta1small}},$ have no closed loops.

\subsection{Three-body, third-order coefficient $\alpha_{3}^{(2)}%
$\label{app:alpha3(33)}}

Continuing to work in the single-particle basis, the contribution
$\raisebox{-0.05in}{\includegraphics[]{beta2small}}$ has the coefficient
\begin{equation}
\alpha_{3}^{(3)}={\sum\limits_{ij\neq00,ik\neq00}}\frac{K_{00;ij}%
K_{j0;0k}K_{ik;00}}{\Delta\varepsilon_{ij}\Delta\varepsilon_{ik}},
\end{equation}
with $i=\{n_{1}l_{1}m_{1}\},$ $j=\{n_{2}l_{2}m_{2}\},$ and $k=\{n_{3}%
l_{3}m_{3}\}.$ Due to angular momentum conservation, we have $l_{1}%
=l_{2}=l_{3}$ and $m_{1}=-m_{2}=-m_{3}.$ Using Eqs.~(\ref{deltaKsp}),
(\ref{Kn1n2l}), and (\ref{EnergyDiffs}), the coefficient is given by%
\begin{align}
&\alpha_{3}^{(3)}=\sum_{n_{1}n_{2}n_{3}l_{1}}^{\omega_{c}/\omega}(2l_{1}+1) \times  \label{alpha33} \\ &\frac
{K_{\text{s.p.}}(n_{1},n_{2},l_{1})K_{\text{s.p.}}(n_{2},n_{3}%
,l_{1})K_{\text{s.p.}}(n_{1},n_{3},l_{1})}{(2n_{1}+2n_{2}+2l_{1}%
)(2n_{1}+2n_{3}+2l_{1})}, \nonumber%
\end{align}
where the sum is over $0<2n_{1}+2n_{2}+2l_{1}<\omega_{c}/\omega$ and
$0<2n_{1}+2n_{3}+2l_{1}<\omega_{c}/\omega.$ The factor $(2l_{1}+1)$ arises due
to the sum over the quantum number $m_{1}$.

\begin{figure}[ptb]
\centering{\includegraphics[width=1\columnwidth]{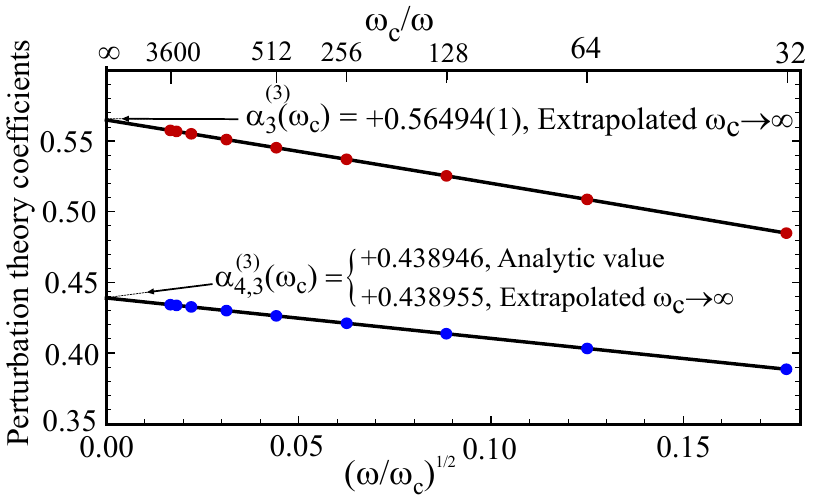}} \caption{(Color online.) Plot of
numerical approximations to the sums which give the coefficients $\alpha
_{3}^{(3)}$ (red circles) and $\alpha_{4,3}^{(3)}$ (blue circles), with a hard
cutoff $\omega_{c}$ in the energy of the intermediate states. The data is
plotted versus $(\omega/\omega_{c})^{1/2}$ on the bottom axis (the top axis
shows the corresponding value of $\omega_{c}/\omega$). The black lines are the
least-square fits to the expected asymptotic behavior $a+b(\omega/\omega
_{c})^{1/2}+c(\omega/\omega_{c})$. The values for $\alpha_{3}^{(3)}$ and
$a_{4,3}^{(3)}$ are obtained by extrapolating to the y-intercept ($\omega
_{c}\rightarrow\infty$). To estimate the one-standard deviation uncertainty in
$\alpha_{3}^{(3)},$ for which we do not have an analytic value, we use the
difference between the extrapolated and analytic values of $\alpha_{4,3}%
^{(3)}$.}%
\label{fig:alpha33}%
\end{figure}

We have not found an analytic expression for $\alpha_{3}^{(3)},$ and the sums
in Eq. (\ref{alpha33}) converge slowly, making precise numerical determination
demanding. We obtain an estimate by fitting numerical approximations versus
$\omega/\omega_{c}$ to the asymptotic form $a+b(\omega/\omega_{c}%
)^{1/2}+c(\omega/\omega_{c}),$ dropping terms that are $\mathcal{O}%
[(\omega/\omega_{c})^{3/2}]$. The best-fit constants $a,b,$ and $c $ give the
curve $\alpha_{3}^{(3)}(\omega_{c})$ shown in Fig.~\ref{fig:alpha33}. The best
estimate for $\alpha_{3}^{(3)}$, found by extrapolating $\omega_{c}%
/\omega\rightarrow\infty$, is%
\begin{equation}
\alpha_{3}^{(3)}=0.56494\pm0.00001.
\end{equation}
To determine the one-standard deviation uncertainty in $\alpha_{3}^{(3)}$
associated with our extrapolation method, we have compared the analytic value
of $\alpha_{4,3}^{(3)}$ given in Eq. (\ref{gamma3}) to the value for
$\alpha_{4,3}^{(3)}$ found by numerical extrapolation. The comparison is shown
in Fig.~\ref{fig:alpha33}.

\subsection{Four-body, third-order coefficients $\alpha_{4,1}^{(3)}$ and
$\alpha_{4,2}^{(3)}$\label{app:alpha41(3)and42(3)}}

The contributions $\raisebox{-0.05in}{\includegraphics[]{gamma1small}}$ and
$\raisebox{-0.05in}{\includegraphics[]{gamma2small}}$ give the coefficients
\begin{equation}
\alpha_{4,1}^{(3)}=\sum\limits_{i\neq0,j}\frac{K_{00;ij}K_{j0;00}K_{i0;00}%
}{\Delta\varepsilon_{ij}\Delta\varepsilon_{i0}}%
\end{equation}
and%
\begin{equation}
\alpha_{4,2}^{(3)}=\sum\limits_{i\neq0,j\neq0}\frac{K_{00;0i}K_{i0;0j}%
K_{j0;00}}{\Delta\varepsilon_{i0}\Delta\varepsilon_{j0}},
\end{equation}
respectively. Due to angular momentum conservation, only $i,j$ with $l=m=0$
contribute. Using Eqs.~(\ref{deltaKsp}), (\ref{Kn1n2l}), and
(\ref{EnergyDiffs}), we obtain the numerical results \begin{widetext}
\begin{align}
\alpha_{4,1}^{(3)}=  \sum_{n_{1}\neq0,n_{2}=0} \frac{K_{\text{s.p.}}%
(n_{1},0,0)K_{\text{s.p.}}(n_{2},0,0)K_{\text{s.p.}}(n_{1},n_{2},0)}%
{4n_{1}(n_{1}+n_{2})}=0.077465...
\end{align}
and%
\begin{align}
\alpha_{4,2}^{(3)}=\sum_{n_{1}\neq0,n_{2}\neq0} \frac{K_{\text{s.p.}}%
(n_{1},0,0)K_{\text{s.p.}}(n_{1},n_{2},0)K_{\text{s.p.}}(n_{2},0,0)}%
{4n_{1}n_{2}}=0.051099....
\end{align}
These are tree-diagram processes, which, like $\alpha_{3}^{(2)},$ converge
quickly, thereby making it is easy to obtain a precise numerical approximation
from a small number of excited orbitals.

\subsection{Four-body, third-order coefficient $\alpha_{4,3}^{(3)}%
$\label{app:alpha43(3)}}

The two-, three-, and four-body contributions
$\raisebox{-0.05in}{\includegraphics[]{alpha3small}},\raisebox{-0.05in}{\includegraphics[]{beta3small}},$
and $\raisebox{-0.05in}{\includegraphics[]{gamma3small}}$ have the same
coefficient,
\begin{align}
\alpha_{4,3}^{(3)}=\sum\limits_{ij\neq00}\frac{K_{00;ij}K_{00;00}K_{ij;00}%
}{\Delta\varepsilon_{ij}^{2}} =\sum\limits_{\tilde{\imath}\tilde{j}\neq00}%
\frac{K_{00;\tilde{\imath}\tilde{j}}K_{00;00}K_{\tilde{\imath}\tilde{j};00}%
}{\Delta\varepsilon_{\tilde{\imath}\tilde{j}}^{2}},
\end{align}
\end{widetext}where in the last expression, rather than evaluating the sums in the
single-particle basis, we observe that $\phi_{000}(\mathbf{r}_{1})\phi
_{000}(\mathbf{r}_{2})=\tilde{\phi}_{000}(\mathbf{r)}\tilde{\Phi}%
_{000}(\mathbf{R)}$ and sum over relative and center-of-mass bases states
$|\tilde{\imath}\tilde{j}\rangle=|nlm,NLM\rangle,$ excluding $\tilde{\imath
}\tilde{j}=00$. The interactions conserve the center-of-mass motion, implying
$\tilde{j}=\{NLM\}=\{000\}$. Angular momentum conservation gives
$\tilde{\imath}=\{n00\}$. Finally, using $\Delta\varepsilon_{n00,000}=2n$ and
$K_{\tilde{\imath}0;00}=K_{n00,000;000,000}=K_{\text{rel}}(n,0)$ from
Eq.~(\ref{Kreln}), we obtain the analytic result%
\begin{align}
\alpha_{4,3}^{(3)}&=\sqrt{\frac{2}{\pi}}\sum_{n>0}\frac{[K_{\text{rel}%
}(n,0)]^{2}}{4n^{2}}\label{gamma3} \\ &=\left(  \frac{2}{\pi}\right)  ^{3/2}[\frac{\pi^{2}}%
{24}+\log2-\frac{1}{2}\left(  \log2\right)  ^{2}]=0.43894....\nonumber %
\end{align}
If we include the exponential regulator, we confirm that $\alpha_{4,3}%
^{(3)}\ $converges as $\ (\omega/\omega_{c})^{1/2}$. Because the sums for
$\alpha_{4,3}^{(3)}$ and $\alpha_{3}^{(3)}$ have the same asymptotic
behaviors, we use the exact result in Eq.~(\ref{gamma3}) to determine the
accuracy of the extrapolation for $\alpha_{3}^{(3)}$ shown in
Fig.~\ref{fig:alpha33}.

\subsection{Five-body, third-order coefficient $\alpha_{5}^{(3)}%
$\label{app:alpha5(3)}}

The three-, four-, and five-body contributions
$\raisebox{-0.1071in}{\includegraphics[]{beta4small}},\raisebox{-0.05in}{\includegraphics[]{gamma4small}},$
and $\raisebox{-0.05in}{\includegraphics[]{delta1small}}$ have the same
coefficient
\begin{equation}
\alpha_{5}^{(3)}=\sum\limits_{i\neq0}\frac{K_{00;0i}K_{00;00}K_{i0;00}}%
{\Delta\varepsilon_{i0}{}^{2}}.
\end{equation}
Working in the single-particle basis $|n_{1}l_{1}m_{1},n_{2}l_{2}m_{2}%
\rangle,$ we obtain the analytic result
\begin{align}
\alpha_{5}^{(3)}  &  =\sqrt{\frac{2}{\pi}}\sum_{n>0}\frac{[K_{\text{s.p.}%
}(n,0)]^{2}}{4n^{2}} \nonumber \\ &=\frac{3}{4(2\pi)^{3/2}}{}_{4}F_{3}%
(1,1,1,5/2;2,2,2;1/4)\nonumber\\
&  =\left(  \frac{2}{\pi}\right)  ^{3/2}[\frac{1}{2}\text{Li}_{2}(1/2-\sqrt
{3}/4)-\log(1+\sqrt{3}/2) \nonumber \\ & \hspace{5mm}-\frac{1}{4}(\log(1+\sqrt{3}/2)-\log2)^{2}%
+\log2]\nonumber\\
&  =0.051916...,
\end{align}
where $_{p}F_{q}$ is a generalized hypergeometric function, and Li$_{2}(z)$ is
the polylogarithm function. Evaluation with a regulator function shows that
this expression converges as $(\omega/\omega_{c})^{1/2}$.

\subsection{Two-body, second-order coefficient $\beta_{2}^{(2)}$%
\label{app:beta2(2)}}

The coefficients $\beta_{m}^{(n)}(\omega)$ diverge when $\omega_{c}%
/\omega\rightarrow\infty$. The two-body contribution
$\raisebox{-0.05in}{\includegraphics[]{alpha1small}}$ has the coefficient
\begin{equation}
\beta_{2}^{(2)}\left(  \omega\right)  =\sum\limits_{ij\neq00}^{\omega
_{c}/\omega}\frac{K_{00;ij}K_{ij;00}}{\Delta\varepsilon_{ij}}=\sum
\limits_{\tilde{\imath}\neq0}^{\omega_{c}/\omega}\frac{K_{00;\tilde{\imath}%
0}K_{\tilde{\imath}0;00}}{\Delta\varepsilon_{\tilde{\imath}0}},
\end{equation}
where we have switched to the relative and center-of-mass basis $|\tilde
{\imath}\tilde{j}\rangle=|nlm,NLM\rangle$ in the last expression. Using the
fact that only $l=m=0$ and $\tilde{j}=0$ states contribute greatly simplifies
the evaluation of $\beta_{2}^{(2)}(\omega)$ by reducing the multidimensional
sum to a single summation. Using $K_{n00,000;000,000}=K_{\text{rel}}(n,0)$
from Eq.~(\ref{Kreln}) and the exponential regulator $1/\Delta\varepsilon
_{\tilde{\imath}0}\rightarrow e^{-2n(\omega/\omega_{c})}/2n,$ we obtain
\begin{align}
&\beta_{2}^{(2)}\left(  \omega\right)  =\sum\limits_{n>0}\frac{[K_{\text{rel}%
}(n,0)]^{2}}{2n}e^{-\frac{2n\omega}{\omega_{c}}} \label{alpha1} \\ &=\left(  \frac{2}{\pi}\right)
\left(  \sqrt{\frac{\omega_{c}}{2\omega}}-(1-\log2)-\frac{3}{2}\sqrt
{\frac{\omega}{2\omega_{c}}}\right)  +\mathcal{O(}1/\omega_{c}).\nonumber
\end{align}
This coefficient diverges as $\sqrt{\omega_{c}/\omega},$ but as shown in the
main body of this paper, the divergence cancels after renormalization, leaving
a finite correction proportional to $(2/\pi)(1-\log2).$

\subsection{Two-body, third-order coefficient $\beta_{2}^{(3)}$%
\label{app:beta2(3)}}

Next we consider the contribution
$\raisebox{-0.05in}{\includegraphics[]{alpha2small}},$ with coefficient \begin{widetext}
\begin{equation}
\beta_{2}^{(3)}\left(  \omega\right)  =\sum\limits_{ij\neq00,kl\neq00}%
\frac{K_{00;ij}K_{ij;kl}K_{kl;00}}{\Delta\varepsilon_{ij}\Delta\varepsilon
_{kl}}=\sum\limits_{\tilde{\imath}\neq0,\tilde{k}\neq0}\frac{K_{00;\tilde
{\imath}0}K_{\tilde{\imath}0;\tilde{k}0}K_{\tilde{k}0;00}}{\Delta
\varepsilon_{\tilde{\imath}0}\Delta\varepsilon_{\tilde{k}0}}, \label{alpha2}%
\end{equation}
where we again switch to relative and center-of-mass basis states and use the
selection rules. Inserting exponential regulators for both energy denominators
in Eq.~(\ref{alpha2}) and using Eq.~(\ref{Krelnn}), it follows that
\begin{align}
\beta_{2}^{(3)}(\omega)  &  =\sum_{n>0,n^{\prime}>0}\frac{K_{\text{rel}%
}(n,0)K_{\text{rel}}(n,n^{\prime})K_{\text{rel}}(n^{\prime},0)}{4nn^{\prime}%
}e^{-\frac{2(n^{\prime}+n)\omega}{\omega_{c}}}\nonumber\\
&  =\sqrt{\frac{\pi}{2}}\left(  \sum_{n>0}\frac{[K_{\text{rel}}(n,0)]^{2}}%
{2n}e^{-\frac{2n\omega}{\omega_{c}}}\right)  \left(  \sum_{n^{\prime}>0}%
\frac{[K_{\text{rel}}(n^{\prime},0)]^{2}}{2n^{\prime}}e^{-\frac{2n^{\prime
}\omega}{\omega}}\right) =[\beta_{2}^{(2)}\left(  \omega\right)  ]^{2}/\alpha_{2}^{(1)}.
\end{align}
This factorization result is important for the renormalization of the two-body
interaction at third- and higher-orders. \end{widetext}

\subsection{Three-body, third-order coefficient $\beta_{3}^{(3)}(\omega
)$\label{app:beta3(3)}}

The contribution $\raisebox{-0.05in}{\includegraphics[]{betactsmall}}$ gives
the coefficient
\begin{align}
\beta_{3}^{(3)}\left(  \omega\right)  &=\sum\limits_{ij\neq00,k\neq0}%
\frac{K_{00;ij}K_{ij;0k}K_{k0;00}}{\Delta\varepsilon_{ij}\Delta\varepsilon
_{k}} \nonumber \\ &=\sum\limits_{\tilde{\imath}\neq0,k\neq0}\frac{K_{00;\tilde{\imath}%
0}K_{\tilde{\imath}0;0k}K_{k0;00}}{\Delta\varepsilon_{\tilde{\imath}0}%
\Delta\varepsilon_{k0}}. \label{betact}%
\end{align}
In the last equality, we replaced the sum over single-particle intermediate
states $|ij\rangle$ with a sum over relative and center-of-mass states
$|\tilde{\imath}\tilde{j}\rangle,$ and then used the selection rule $\tilde
{j}=0.$ The sum over $k$ remains over the single-particle basis. We therefore
require the \textquotedblleft mixed-basis\textquotedblright\ matrix elements
$K_{\tilde{\imath}0;0k}.$ Using the selection rules $l=m=0$ for the relative
motion, and $l_{1}=m_{1}=0$ for the single-particle motion, we need only
$K_{\tilde{\imath}0;0k}=K_{\text{mixed}}(n,n_{1})$ with $\tilde{\imath
}=\{n00\}$ and $k=\{n_{1}00\},$ where \begin{widetext}
\begin{align}
K_{\text{mixed}}(n,n_{1})  &=\sqrt{\frac{2}{\pi}}(2\pi)^{3/2}  \int\tilde
{\phi}_{n00}^{\ast}\left(  \mathbf{r}\right)  \tilde{\phi}_{000}^{\ast
}(\mathbf{R})\delta^{(3)}(\mathbf{r)}\phi_{n_{1}00}(\mathbf{r}_{1})\phi
_{000}(\mathbf{r}_{2})d\mathbf{r}d\mathbf{R}\nonumber\\
&  =\sqrt{\frac{2}{\pi}}(2\pi)^{3/2}\tilde{\phi}_{n00}\left(  \mathbf{0}%
\right)  \int\tilde{\phi}_{000}^{\ast}(\mathbf{R})\phi_{n_{1}00}%
(\frac{\mathbf{R}}{\sqrt{2}})\phi_{000}(\frac{\mathbf{R}}{\sqrt{2}%
})d\mathbf{R,}%
\end{align}
and $\mathbf{r}_{1,2}=(\mathbf{R}\pm\mathbf{r)/}\sqrt{2}$. Substituting in
harmonic oscillator wavefunctions gives%
\begin{align}
K_{\text{mixed}}(n,n_{1})=16\sqrt{2}\tilde{\phi}_{n00}\left(  \mathbf{0}%
\right)  N_{n_{1}0}\int_{0}^{\infty}L_{n_{1}}^{(1/2)}\left(  x^{2}\right)
e^{-2x^{2}}x^{2}dx,
\end{align}
where $x=\left\vert \mathbf{R}\right\vert /\sqrt{2}$ and we have integrated
over the angles. Noting that the remaining integral over $x$ is proportional
to $K_{\text{s.p.}}(n_{1},0,0)$ in Eq.~(\ref{Kspintegral}), we find that
\begin{equation}
K_{\text{mixed}}(n,n_{1})=\sqrt{\frac{\pi}{2}}K_{\text{rel}}%
(n,0)K_{\text{s.p.}}(n_{1},0,0).
\end{equation}
Inserting exponential regulators for each energy denominator in Eq.
(\ref{betact}), we obtain 
\begin{align}
\beta_{3}^{(3)}(\omega)   &=\sum_{n>0,n_{1}>0}\frac{K_{\text{rel}%
}(n,0)K_{\text{mixed}}(n,n_{1})K_{\text{s.p.}}(n_{1},0,0)}{4nn_{1}} e^{-\frac{2(n+n_{1})\omega}{\omega_{c}}}\nonumber\\
&  =\sqrt{\frac{\pi}{2}}\left(  \sum_{n_{1}>0}\frac{[K_{\text{s.p.}}%
(n_{1},0,0)]^{2}}{2n_{1}}e^{-\frac{2n_{1}\omega}{\omega_{c}}}\right) \left(
\sum_{n>0}\frac{[K_{\text{rel}}(n,0)]^{2}}{2n}e^{-\frac{2n\omega}{\omega_{c}}%
}\right)  =\alpha_{3}^{(2)}\beta_{2}^{(2)}(\omega)/\alpha_{2}^{(1)}.
\end{align}
The factorization of $\beta_{3}^{(3)}$ in the finite part $\alpha_{3}%
^{(2)}/\alpha_{2}^{(1)}$ and the divergent part $\beta_{2}^{(2)}(\omega)$ is
important for the renormalization of the three-body interaction at third order. \end{widetext}

\bibliography{refs}

\end{document}